# Probing photo-induced granular superconductivity in $K_3C_{60}$ thin films with an ultrafast on-chip voltmeter

J. D. Adelinia[1†], E. Wang[1†], M. Chavez-Cervantes[1], T. Matsuyama[1], M. Fechner[1], M. Buzzi[1], G. Meier[1], A. Cavalleri[1,2]

[1] *Max Planck Institute for the Structure and Dynamics of Matter, Hamburg, Germany*
[2] *Department of Physics, Clarendon Laboratory, University of Oxford, Oxford, United Kingdom*
† *These authors contributed equally to this work*

**The physics of optically-induced superconductivity remains poorly understood, with questions that range from the underlying microscopic mechanism to the macroscopic electrical response of the non-equilibrium phase. In this paper, we study optically-induced superconductivity in $K_3C_{60}$ thin films, which display signatures of granularity both in the equilibrium state below $T_c$ and in the non-equilibrium photo-induced phase above $T_c$. Photo-conductive switches are used to measure the ultrafast voltage drop across a $K_3C_{60}$ film as a function of time after irradiation, both below and above $T_c$. These measurements reveal fast changes associated with the kinetic inductance of in-grain superconductivity, and a slower response attributed to the Josephson dynamics at the weak links. Fits to the data yield estimates of the in-grain photo-induced superfluid density after the drive and the dynamics of phase slips at the weak links. This work underscores the increasing ability to make electrical measurements at ultrafast speeds in optically-driven quantum materials, and demonstrates a striking new platform for optoelectronic device applications.**

Optical excitation at mid-infrared and terahertz frequencies has emerged as a powerful technique for controlling non-equilibrium functional phenomena in quantum materials[1], including magnetism[2,3], transient topological phases[4,5], and photo-induced superconductivity[6-16].

In the case of optically-manipulated superconductivity, a number of experiments have focused on the alkali-doped fulleride $K_3C_{60}$ (Fig. 1a). Transient superconducting-like optical features have been observed for mid[8-10]- and far[12]-infrared excitation at base temperatures far above the equilibrium superconducting transition temperature $T_c$. In these experiments, which were performed in pressed powder pellets, a long-lived phase with vanishing electrical resistance has been reported[10], further underscoring the similarities with an equilibrium superconductor.

These experiments have recently been complemented by a study in $K_3C_{60}$ thin films[11], which were integrated into an optoelectronic platform[5,11,17-20] to study the transport properties at picosecond timescales. New evidence for a photo-induced superconducting-like response was obtained from these measurements, including those arising from characteristic nonlinear current-voltage behaviour that is absent in the metallic phase. A key feature of the non-equilibrium response was that the photo-induced phase had a non-linear current response with vanishing critical current, an effect that is also observed in equilibrium for temperatures in the range $0.6T_c < T < T_c$, and associated with a granular response. The response in these films was found to be smaller and to survive up to lower temperature than previously reported for powder samples[8-10,12].

Here, we report measurements of the voltage dynamics after photo-excitation in the presence of a quasi-DC current bias, enabling a systematic study of non-equilibrium granular superconductivity. An optoelectronic device incorporating a $K_3C_{60}$ thin film was

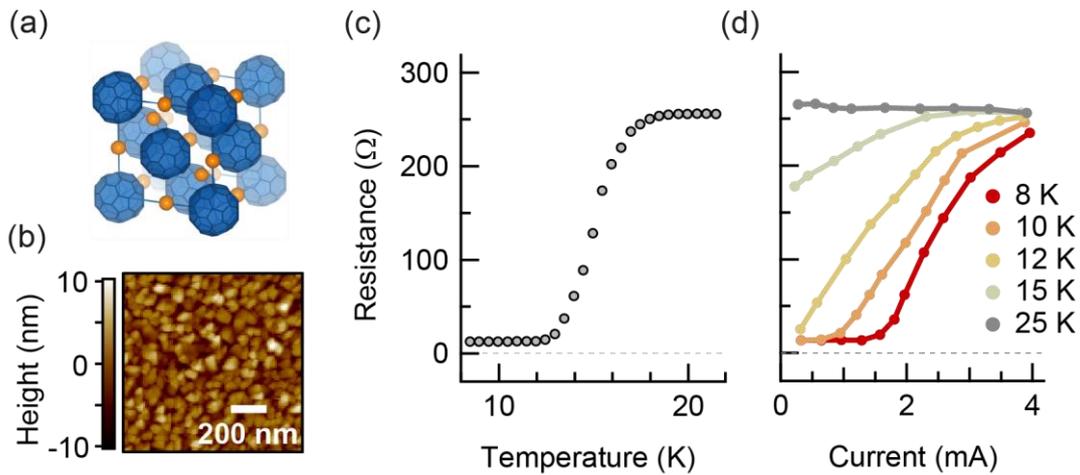

**Figure 1 | Crystal structure and equilibrium transport properties of thin-film $K_3C_{60}$.** **a**, Unit cell of $K_3C_{60}$. $C_{60}$ molecules are represented in blue, and are situated at the lattice points of a face-centred cubic lattice. Potassium atoms are represented in orange. **b**, Atomic force microscopy image of the MBE-grown $C_{60}$ thin film before doping with Potassium. Grains of approximately 100-nm size are visible. **c**, Two-contact DC measurement of resistance vs temperature of the $K_3C_{60}$ thin film. **d**, Two-contact quasi-DC measurement of resistance vs current, for the temperatures 8 K, 10 K, 12 K, 15 K, and 25 K.

---

used to detect ultrafast voltage changes after photo-excitation. These measurements were performed both for base temperatures $T < T_c$, for which the equilibrium superconductor is disrupted or strongly weakened by mid-infrared excitation, and for base temperatures $T > T_c$, for which a photo-induced superconducting-like response is observed.

Thin films of $K_3C_{60}$ were grown on sapphire substrates via molecular-beam epitaxy. An atomic force micrograph of the sample surface is shown in Fig. 1b. The resistance of the $K_3C_{60}$ thin film vs temperature was measured in DC in a two-contact configuration (Fig. 1c), and exhibits a broad superconducting transition with $T_c$ = 19 K, which reflects a granular response in the temperature range 12 K < T < 19 K. The current-dependence of the resistance, as measured under a quasi-DC bias [see supplementary], is shown in Fig. 1d. Nonlinear current-voltage characteristics, typical for superconductivity, were seen at temperatures below $T_c$.

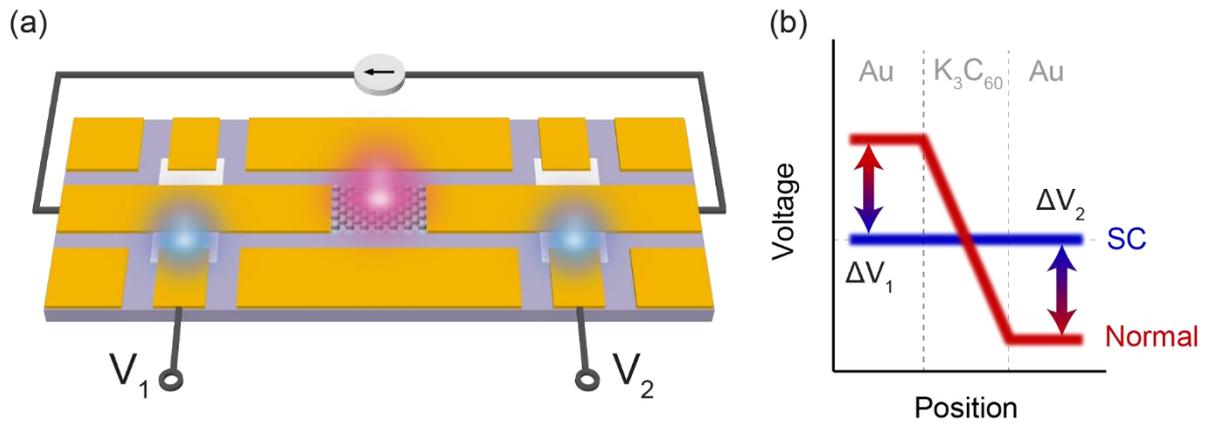

**Figure 2 | Ultrafast voltmeter.  a**, Device architecture. An MBE-grown $K_3C_{60}$ thin film was incorporated into a coplanar waveguide, and biased with a quasi-DC current. The $K_3C_{60}$ was then excited with a 300-fs duration, 7-μm wavelength laser pulse at a fluence of 5 mJ/cm² (indicated in red). The resulting perturbation to the voltage drop across the $K_3C_{60}$ thin film generated two electrical pulses which propagated left and right along the coplanar waveguide. The electrical pulses were detected on the left and right using silicon photo-conductive switches, which were operated with 250-fs gating pulses at 515-nm wavelength (indicated in cyan). **b**, Sketch of the voltage across the device. The blue (red) line represents the voltage profile in the superconducting (normal) state of $K_3C_{60}$. The arrows represent the voltage changes at the positions of the photo-conductive switches during an ultrafast phase transition.

---

The device architecture for the ultrafast transport measurements is illustrated in Fig. 2a. The $K_3C_{60}$ samples were connected to the signal line of a coplanar waveguide, which allowed for voltage transients to propagate at frequencies of up to 1 THz.[11,17,19] The coplanar waveguide was contacted to photo-conductive switches[21] on both sides of the sample, which were used to detect the transient voltages. The switches consisted of patches of amorphous silicon, which becomes conductive when excited with an ultrafast laser pulse in the visible spectrum[5,21]. When photo-excited under a voltage bias, the switches launched current pulses with duration determined by the lifetime of the photo-excited carriers (< 1 ps) and a total charge proportional to the voltage bias, thus enabling sampling of transient voltages. By varying the time delay between the $K_3C_{60}$ excitation pulse (pump) and the switch trigger pulse (probe), a quantitative voltage measurement was obtained with sub-picosecond time resolution.

The K$_3$C$_{60}$ thin film was biased with a voltage pulse of 500-ns duration, long enough for the local current to stabilise (see supplementary Fig. S6). After a constant current flow was established, the K$_3$C$_{60}$ was excited with a laser pulse of 300-fs duration, 7-μm wavelength, and 5-mJ/cm$^2$ excitation fluence. In-current photo-excitation of the K$_3$C$_{60}$ caused changes in the voltage drop across the sample (Fig. 2b). On the ultrafast timescales probed in the measurement, the voltage in the device was effectively floating, so changes in voltage across the K$_3$C$_{60}$ thin film resulted in equal and opposite voltage pulses propagating in the left and right directions. The voltage pulses were then detected at the photo-conductive switches as transient variations in the voltages V$_1$ and V$_2$. The switches were operated with laser pulses of 250-fs duration at 515-nm wavelength.

The first set of measurements was carried out at temperatures below T$_c$. In this regime, the photon energy of the excitation pulse was larger than the energy gap of the equilibrium superconducting state in K$_3$C$_{60}$.[22] In this regime, mid-infrared excitation is expected to break Cooper pairs into unpaired electrons and disrupt superconductivity[23], either partially or completely depending on the pump fluence. Hence, for this type of excitation of the equilibrium superconducting state, an increase in the voltage drop across the sample is expected[24,25] (Fig. 3a).

The measured pump-induced changes in V$_1$ and V$_2$ (henceforth referred to as ΔV$_1$ and ΔV$_2$) are plotted in Fig. 3b as a function of pump-probe delay. All voltages are normalised by a factor $V_0$, which, for the measurements below T$_c$, is defined as the applied current multiplied by the sample resistance at 20 K. ΔV$_1$ exhibited a step-like increase with a rise time of approximately 1 ps. ΔV$_2$ displayed equal and opposite behaviour to ΔV$_1$. For the given direction of applied current, this indicates an increase in the total voltage drop across the sample, consistent with disruption of superconductivity. The total pump-induced change in voltage drop can be calculated as ΔV$_1$ − ΔV$_2$. Fig. 3c displays the pump-

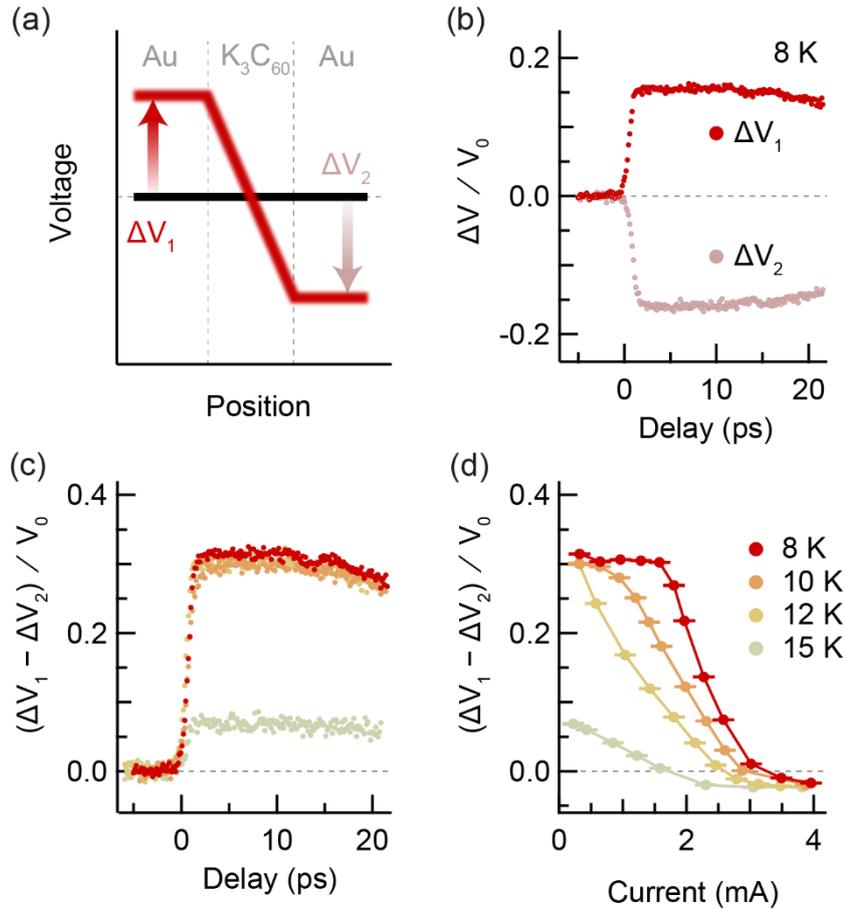

**Figure 3 | Photo-induced voltage dynamics in $K_3C_{60}$ (T < $T_c$). a**, Sketch of the voltage across the device. The voltage profile changes from the equilibrium superconducting state (black) to the disrupted state (red) upon photo-excitation. **b**, Pump-induced change in voltage on left and right of sample vs pump-gate delay, measured at 8 K. **c**, Pump-induced change in voltage drop across sample vs pump-gate delay, measured at 8 K, 10 K, 12 K, and 15 K, with an applied current of 0.3 mA. **d**, Pump-induced change in voltage drop across sample vs applied current, measured at the same temperatures as in (c). Each data point is the mean value of the time-trace within a 6-ps window centred at 6-ps delay. The error bars represent the standard error of the mean change in voltage drop. All voltages are normalized by a factor $V_0$, which is defined for the measurements below $T_c$ as the applied current multiplied by the sample resistance at 20 K.

---

induced change in voltage drop across the $K_3C_{60}$ thin film as a function of pump-probe delay for different temperatures below $T_c$. Fig. 3d displays the same normalised change in voltage drop as a function of applied current. Each data point in the current dependence is the mean value of the time-trace within a 6-ps window centred at 6-ps delay. At 8 K, the pump-probe signal was constant with the applied current up until 2 mA, above which the signal exhibited a step-like decrease. The step was consistent with the critical current of the equilibrium superconducting state (Fig. 1d). Upon increasing temperature, the pump-probe signal began to decrease at lower currents and the step broadened.

The reduction in pump-probe signal with increasing temperature and current are consistent with the onset of resistivity in the equilibrium superconducting state before excitation. This is well-understood quantitatively when comparing these results to a circuit model to simulate the voltage dynamics. Two factors govern the physics of this model. First, disruption of superconductivity involves a transformation of charge carriers from Cooper pairs to resistive quasiparticles[23]. Thus, the carrier dynamics should be well-captured by using a two-fluid model consisting of a resistive channel in parallel with a purely inductive channel. Second, the granular constitution of the $K_3C_{60}$ thin film impedes dissipationless superconducting transport at temperatures just below the bulk $T_c$, as evidenced in equilibrium by the broadened superconducting transition (Fig. 1d). This is understood to result from thermally-activated phase slips between grain boundaries, and can be modelled using a resistively- and capacitively-shunted Josephson junction (RCSJ) model[26]. We therefore incorporate the RCSJ model into the superconducting channel of a two-fluid model, forming a granular two-fluid model, in order to simulate the voltage dynamics in a granular superconductor (Fig. 4b).

The voltage dynamics for the disruption of superconductivity under a DC current bias were simulated by numerically solving the circuit equations for the granular two-fluid model under a changing superfluid density (more details in supplementary material). The effect of temperature was modelled by considering thermal fluctuations in the phases of the order parameters of the superconducting grains, which result in thermally-activated phase slips and the onset of resistivity at temperatures close to $T_c$. We have assumed that, before time-zero, the superfluid density $n_s$ depends on temperature $T$ following the relation[27]

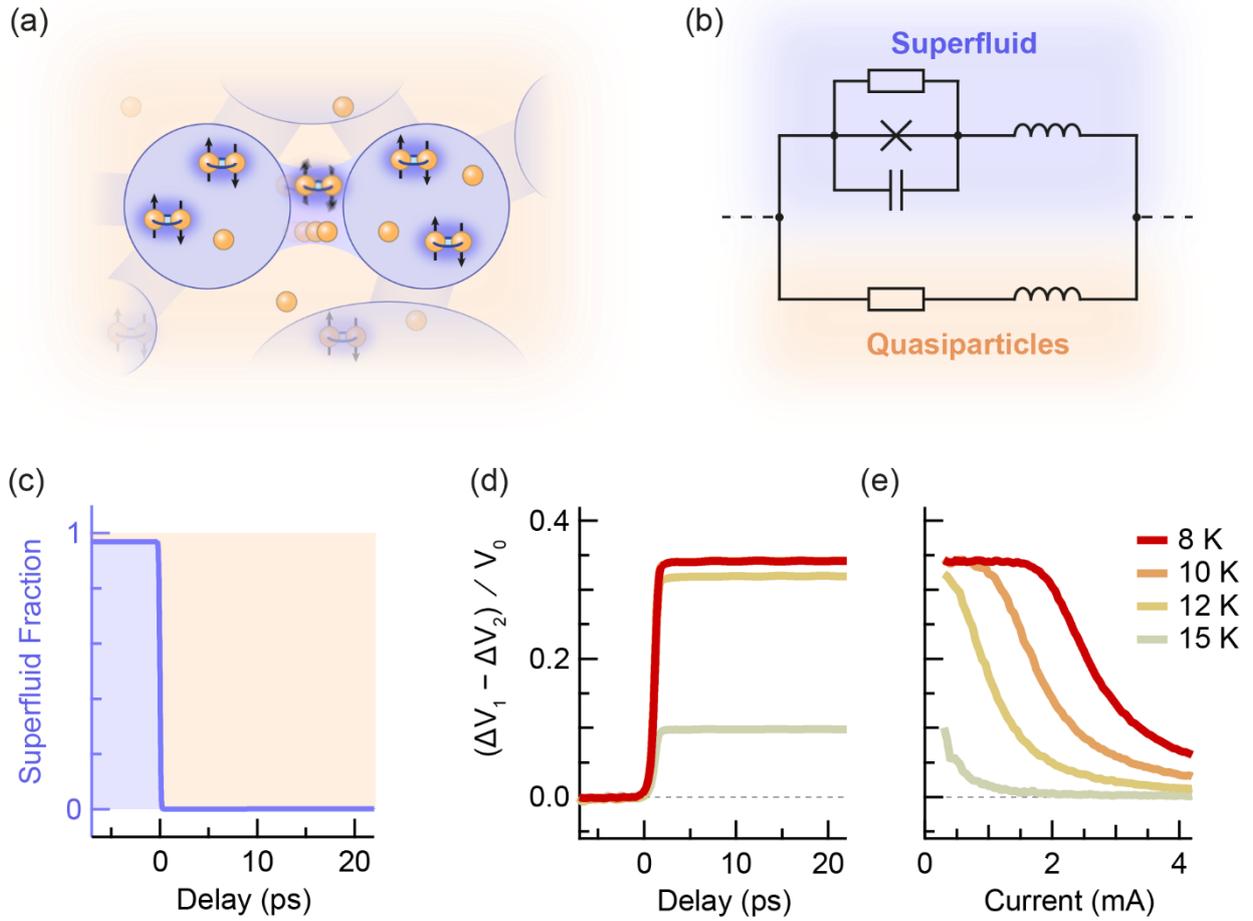

**Figure 4 | Disruption of superconductivity with a granular two-fluid model. a**, Sketch of a granular material with Cooper pairs and quasiparticles. Cooper pairs tunnel across grain boundaries, whilst thermal phase fluctuations disrupt phase coupling between the different superconducting grains. **b**, Circuit diagram of the granular two-fluid model used to represent the $K_3C_{60}$ sample. The superfluid channel (highlighted in purple) consists of a typical RCSJ model extended to include the kinetic inductance of the superconducting grains. The resistive quasiparticle channel is highlighted in orange. **c**, Time-dependent superfluid density used in the disruption simulation at 8 K. The superfluid fraction is assumed to reduce to zero at 0-ps delay. **d**, Simulated change in voltage drop across sample vs delay, for base temperatures of 8 K, 10 K, 12 K, and 15 K, with an applied current of 0.3 mA. **e**, Simulated change in voltage drop across sample vs applied current, for the same base temperatures as in (d). All voltages are normalized by a factor $V_0$, which is defined for the measurements below $T_c$ as the applied current multiplied by the sample resistance at 20 K.

$$n_s \propto 1 - (T/T_c)^4. \tag{1}$$

At time-zero, the superfluid density was assumed to reduce from $n_s(T)$ to zero, as shown in Fig. 4c. The calculated voltage transients are displayed in Fig. 4d and Fig. 4e as functions of delay and current, respectively. The results from the granular two-fluid model exhibit strong agreement with the experimental results in Fig. 3.

In a second set of measurements, we examined the photo-induced non-equilibrium state at temperatures above $T_c$. The experimental setup was identical to the previous measurements below $T_c$. Before photo-excitation, a voltage drop was present across the metallic $K_3C_{60}$ thin film corresponding to its equilibrium state resistance and the applied bias current. In this case, for the conditions in which non-equilibrium superconductivity is induced, a decrease in voltage drop upon photo-excitation is expected (Fig. 5a).

The measured $\Delta V_1$ and $\Delta V_2$ for photo-excitation at 25 K are plotted in Fig. 5b as a function of pump-probe delay. All voltages are normalised by a factor $V_0$, which, for the measurements above $T_c$, is defined as the applied bias multiplied by the sample resistance at the measurement temperature. The signs of $\Delta V_1$ and $\Delta V_2$ were reversed compared to the measurements below $T_c$, indicating a reduction in voltage drop upon photo-excitation and an increase in the conductivity of the sample. $\Delta V_1$ became negative after photo-excitation, with dynamics present on two timescales; a fast, negative spike was present immediately after time-zero, which was followed by a slow decrease of ~10-ps duration. $\Delta V_2$ displayed equal and opposite behaviour to $\Delta V_1$. Fig. 5c displays the pump-induced change in voltage drop across the $K_3C_{60}$ thin film as a function of pump-probe delay for different temperatures above $T_c$. The magnitude of the voltage change decreased with increasing temperature, until the sign of the response was reversed at 80 K. The positive response at high temperatures was also qualitatively different; only a single rise and decay was observed.

Fig. 5d displays the voltage change as a function of the applied current. Each data point in the current dependence is the mean value of the time-trace within a 5-ps window centred at 11-ps delay. At 25 K, linear behaviour in current was observed for small currents. Above

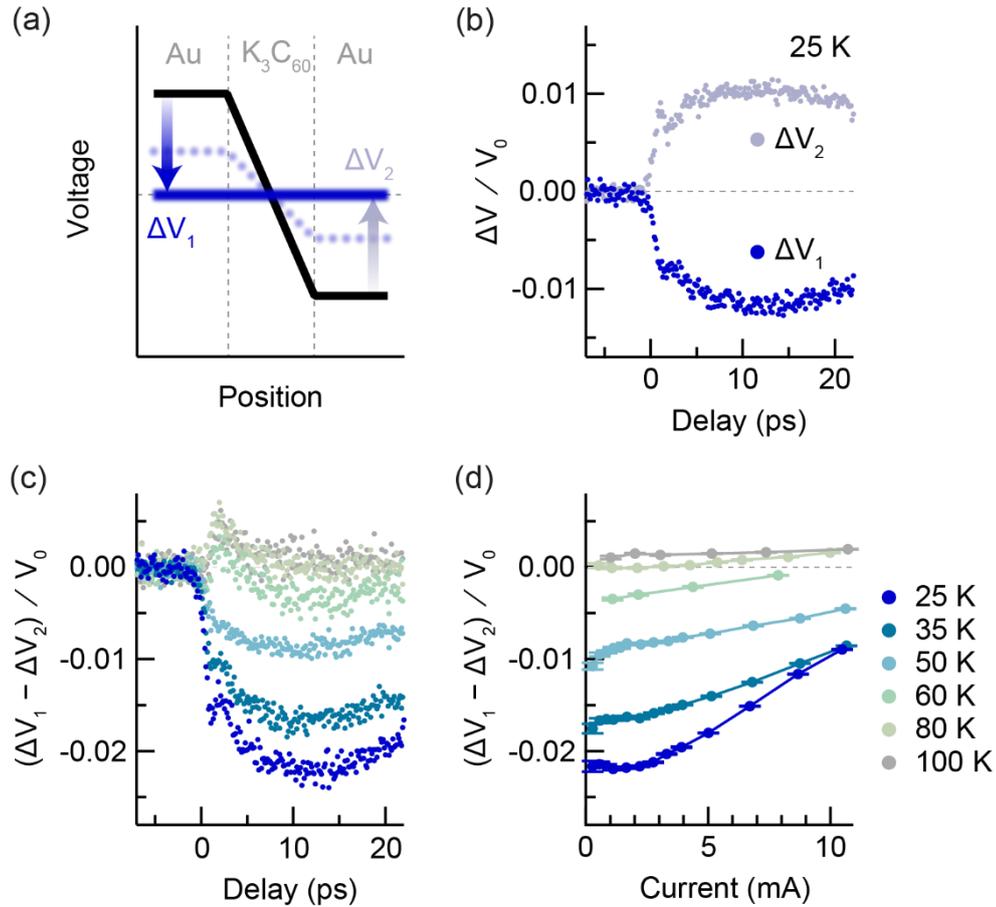

**Figure 5 | Photo-induced voltage dynamics in $K_3C_{60}$ (T > $T_c$).** **a**, Sketch of the voltage across the device. The voltage profile changes when the sample is photo-excited from the equilibrium metallic state (black) into the nonequilibrium superconducting-like state (blue). The dotted blue line represents a superconducting-like state that has finite resistivity due to the effects of granularity. **b**, Pump-induced change in voltage on left and right of sample vs pump-gate delay, measured at 25 K. **c**, Pump-induced change in voltage drop across sample vs pump-gate delay, measured at 25 K, 35 K, 50 K, 60 K, 80 K, and 100 K, with an applied current of 1 mA. **d**, Pump-induced change in voltage drop across sample vs applied current, measured at the same temperatures as in (c). Nonlinear I-V behaviour is observable, which is also suppressed with increasing temperature. Each data point is the mean value of the respective time-trace within a 5-ps window centred at 11-ps delay. The error bars represent the standard error of the mean. All voltages are normalized by a factor $V_0$, which is defined for the measurements above $T_c$ as the applied current multiplied by the equilibrium sample resistance at the measurement temperature.

---

2 mA, the signal became nonlinear with the applied current. Upon increasing the temperature, the onset of the nonlinear behaviour was shifted to lower currents. The nonlinear effect was also reduced alongside the signal magnitude at higher temperatures;

the positive voltage change at 100 K was linear in current across the whole measurement range.

To interpret the experimental data above $T_c$, we consider the two-fluid dynamics of the charge carriers upon photo-excitation, in addition to the previously-discussed role of granularity in the photo-induced state.[11] We attribute the fast response in the data to a change in the kinetic inductance of the $K_3C_{60}$ thin film. The kinetic inductance $L_K$ of a superconductor is given by[25]

$$L_K = \frac{\alpha}{n_s}, \qquad (2)$$

where $\alpha$ is a constant that depends on the material geometry and the effective mass of the charge carriers. A change in the superfluid density will therefore change the kinetic inductance. When this occurs in an applied current $I$, this generates an additional voltage

$$V_K = I\frac{dL_K}{dt} \propto -\frac{I}{n_s^2}\frac{dn_s}{dt}. \qquad (3)$$

For an increase in superfluid density, this voltage is negative, which is consistent with the measurement.

We speculate that the subsequent slow dynamics result from the stabilisation of thermal phase fluctuations. In the case that the phase is highly disordered upon photo-excitation, the initial resistivity of the photo-excited state at early delays would be left unchanged from that of the metallic state. As the phase thermalizes, the rate of thermally-induced phase slips decreases, and the resistivity drops on a timescale of several picoseconds. Nevertheless, the phase slip rate at longer delays would be significant due to lattice heating, resulting in finite resistivity in the long-lived state after photo-excitation.

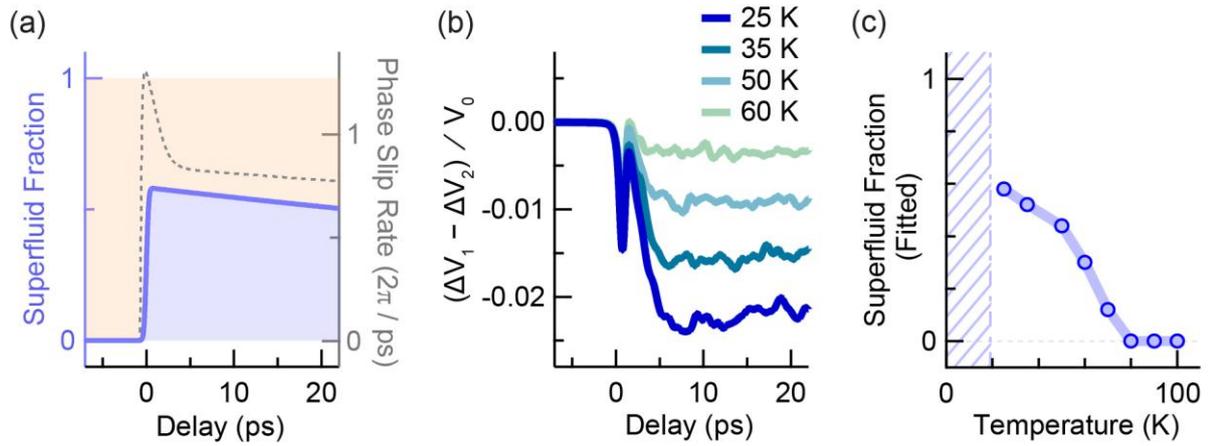

**Figure 6 | Photo-induced superconductivity with a granular two-fluid model. a**, Time profile of superfluid fraction for the simulation of the photo-excited state. The superfluid fraction (purple) is assumed to increase from zero to a finite value at 0-ps delay, and then to decay exponentially with a time constant of 150 ps. The thermally induced phase slip rate (dashed grey line) relates to the thermal energy imparted by the pump pulse. It is large at 0-ps delay and decays to match the lattice temperature at longer delays. **b**, Simulated change in voltage drop across sample vs delay, using peak superfluid fraction as the fitting parameter, for base temperatures of 25 K, 35 K, 50 K, and 60 K, with an applied current of 1 mA. All voltages are normalized by a factor $V_0$, which is defined for the measurements above $T_c$ as the applied current multiplied by the equilibrium sample resistance at the measurement temperature. **c**, Light-induced 'superfluid fraction' vs temperature as obtained from the fitting procedure described in the text. The blue symbols indicate granular two-fluid model fits to the voltage data on the granular thin films. The purple shaded region indicates temperatures below the equilibrium $T_c$.

In order to model the dynamical system, we extend our treatment of the photo-induced state to the previously-discussed granular two-fluid model. The superfluid fraction (Fig. 6a, solid line) was assumed to increase from zero to a finite value at time-zero, with a rise time of 300 fs, and to subsequently decay exponentially in time with a time constant of 150 ps (see supplementary, fitting from lifetime data of Ref. 11). The Josephson phase was assumed to start highly disordered and strongly fluctuating after photo-excitation. Then, the thermal phase fluctuations reduce within the first few picoseconds as the electrons thermalize with the lattice[28,29]. The electron and lattice temperatures after photo-excitation were determined by the mid-infrared pump pulse energy absorbed by the $K_3C_{60}$ sample[30-32] (see supplementary for details). The rate of thermally-induced phase slips is estimated in Fig. 6a (dashed line) by considering thermal hopping over a washboard potential. The circuit simulation at each temperature was used to fit the slow response in the data at 10-ps delay, using the peak superfluid fraction at time-zero as a fitting parameter. Other parameters, such as the rise and decay times for the superfluid fraction, were left unchanged between temperatures.

The fits are displayed in Fig. 6b. The two-timescale dynamics seen in the measurement are also present in the circuit calculations, with a fast, negative voltage spike appearing at time-zero, followed by a slower dip in voltage. The superfluid fraction fitting parameter is plotted in Fig. 6c as a function of temperature, and decreases with increasing temperature. For temperatures of 80 K and above, the experimental data no longer display a reduction in resistivity and are represented with a superfluid fraction of zero in place of a fit. One possible reason for the difference between the temperature scaling reported here and previous reflectivity measurements[8-10] is the fact that lower frequencies are probed in transport, where thermally-induced phase-slip effects may be more pronounced.

The nonlinear I-V characteristics in this measurement (Fig. 5d) are qualitatively different to those reported for measurements with picosecond current pulses.[11] For the data reported here, the current is applied before photo-excitation. The measurement therefore does not distinguish between the effect of the current on the non-equilibrium state and suppression of the photo-susceptibility due to the current. The signal from the non-equilibrium phase at 25 K persists at larger currents than the signal from the equilibrium superconducting state. Furthermore, for photo-excitation below $T_c$, we observe a reversal in sign of the signal for applied bias currents above the critical current (Fig. 3d), which may indicate a revival of superconductivity under large applied currents.

In summary, we report ultrafast voltage measurements on $K_3C_{60}$ thin films upon in-current photo-excitation, and characterized the response of the system for both the photo-induced disruption of the equilibrium low-temperature superconductor, as well as the regime in which a photo-induced superconducting state is achieved. We report a reduction in voltage drop when exciting into the non-equilibrium state above $T_c$. A two-fluid model based on a granular superconductor was used to interpret these findings. The data are consistent with a change in the kinetic inductance upon photo-excitation, followed by a slow reduction in resistivity which can be attributed to thermalization of phase fluctuations. The new understanding extracted from these data complements that obtained for the case of cuprates[6,7], for charge transfer salts[14,15], and for $K_3C_{60}$ powders[8-10,12], for which the responses were compatible with optically-induced bulk superconductivity. The ultrafast voltmeter platform developed here can be extended to other non-equilibrium superconductors and is crucial to understanding inhomogeneities and quenched dynamics in these systems.

**References**

1  Basov, D. N., Averitt, R. D. & Hsieh, D. Towards properties on demand in quantum materials. *Nature Materials* **16**, 1077-1088 (2017).
2  Radu, I. *et al.* Transient ferromagnetic-like state mediating ultrafast reversal of antiferromagnetically coupled spins. *Nature* **472**, 205-208 (2011).
3  Disa, A. S. *et al.* Photo-induced high-temperature ferromagnetism in YTiO$_3$. *Nature* **617**, 73-78 (2023).
4  Wang, Y. H., Steinberg, H., Jarillo-Herrero, P. & Gedik, N. Observation of Floquet-Bloch States on the Surface of a Topological Insulator. *Science* **342**, 453-457 (2013).
5  McIver, J. W. *et al.* Light-induced anomalous Hall effect in graphene. *Nat Phys* **16**, 38-41 (2020).
6  Fausti, D. *et al.* Light-induced superconductivity in a stripe-ordered cuprate. *Science* **331**, 189-191 (2011).
7  Hu, W. *et al.* Optically enhanced coherent transport in YBa$_2$Cu$_3$O$_{6.5}$ by ultrafast redistribution of interlayer coupling. *Nature Materials* **13**, 705-711 (2014).
8  Mitrano, M. *et al.* Possible light-induced superconductivity in K$_3$C$_{60}$ at high temperature. *Nature* **530**, 461-464 (2016).

9   Cantaluppi, A. *et al.* Pressure tuning of light-induced superconductivity in $K_3C_{60}$. *Nature Physics* **14**, 837-841 (2018).
10  Budden, M. *et al.* Evidence for metastable photo-induced superconductivity in $K_3C_{60}$. *Nature Physics* **17**, 611-618 (2021).
11  Wang, E. *et al.* Superconducting nonlinear transport in optically driven high-temperature $K_3C_{60}$. *Nature Communications* **14**, 7233 (2023).
12  Rowe, E. *et al.* Resonant enhancement of photo-induced superconductivity in $K_3C_{60}$. *Nature Physics* (2023).
13  Cremin, K. A. *et al.* Photoenhanced metastable c-axis electrodynamics in stripe-ordered cuprate $La_{1.885}Ba_{0.115}CuO_4$. *Proceedings of the National Academy of Sciences* **116**, 19875-19879 (2019).
14  Buzzi, M. *et al.* Photomolecular High-Temperature Superconductivity. *Physical Review X* **10**, 031028-031028 (2020).
15  Buzzi, M. *et al.* Phase Diagram for Light-Induced Superconductivity in κ-(ET)$_2$-X. *Physical Review Letters* **127**, 197002 (2021).
16  Isoyama, K. *et al.* Light-induced enhancement of superconductivity in iron-based superconductor $FeSe_{0.5}Te_{0.5}$. *Communications Physics* **4**, 160 (2021).
17  Wood, C. D. *et al.* On-chip terahertz spectroscopic techniques for measuring mesoscopic quantum systems. *Review of Scientific Instruments* **84** (2013).
18  Gallagher, P. *et al.* Quantum-critical conductivity of the Dirac fluid in graphene. *Science* **364**, 158-162 (2019).
19  Yoshioka, K., Kumada, N., Muraki, K. & Hashisaka, M. On-chip coherent frequency-domain THz spectroscopy for electrical transport. *Applied Physics Letters* **117** (2020).
20  Potts, A. M. *et al.* On-Chip Time-Domain Terahertz Spectroscopy of Superconducting Films below the Diffraction Limit. *Nano Letters* **23**, 3835-3841 (2023).
21  Auston, D. H. Picosecond optoelectronic switching and gating in silicon. *Applied Physics Letters* **26**, 101-103 (1975).
22  Degiorgi, L., Briceno, G., Fuhrer, M. S., Zettl, A. & Wachter, P. Optical measurements of the superconducting gap in single-crystal $K_3C_{60}$ and $Rb_3C_{60}$. *Nature* **369**, 541-543 (1994).
23  Owen, C. S. & Scalapino, D. J. Superconducting State under the Influence of External Dynamic Pair Breaking. *Physical Review Letters* **28**, 1559-1561 (1972).
24  Testardi, L. R. Destruction of Superconductivity by Laser Light. *Physical Review B* **4**, 2189-2196 (1971).
25  Heusinger, M. A., Semenov, A. D., Nebosis, R. S., Gousev, Y. P. & Renk, K. F. Nonthermal Kinetic Inductance Photoresponse of Thin Superconducting Films. *IEEE Transactions on Applied Superconductivity* **5**, 2595-2598 (1995).
26  Ambegaokar, V. & Halperin, B. I. Voltage Due to Thermal Noise in the DC Josephson Effect. *Physical Review Letters* **22**, 1364-1366 (1969).
27  Tinkham, M. *Introduction to Superconductivity, 2nd Edition*.  (Dover Publications, Inc., 2004).
28  Elsayed-Ali, H. E., Norris, T. B., Pessot, M. A. & Mourou, G. A. Time-resolved observation of electron-phonon relaxation in copper. *Physical Review Letters* **58**, 1212-1215 (1987).
29  Sun, C. K., Vallée, F., Acioli, L., Ippen, E. P. & Fujimoto, J. G. Femtosecond investigation of electron thermalization in gold. *Physical Review B* **48**, 12365-12368 (1993).

30  Semenov, A. D., Nebosis, R. S., Gousev, Y. P., Heusinger, M. A. & Renk, K. F. Analysis of the nonequilibrium photoresponse of superconducting films to pulsed radiation by use of a two-temperature model. *Physical Review B* **52**, 581-581 (1995).
31  Sobolewski, R. Ultrafast dynamics of nonequilibrium quasi-particles in high-temperature superconductors. *SPIE* **3481**, 480-491 (1998).
32  Allen, K. & Hellman, F. Specific heat of $C_{60}$ and $K_3C_{60}$ thin films for T=6—400 K. *Physical Review B* **60**, 11765-11772 (1999).

**Acknowledgements**

The research leading to these results received funding from the European Research Council under the European Union's Seventh Framework Program (FP7/2007-2013)/ERC Grant Agreement No. 319286 (QMAC, A.C.). We acknowledge support from the Deutsche Forschungsgemeinschaft (DFG) via the Cluster of Excellence 'CUI: Advanced Imaging of Matter' - EXC 2056 - project ID 390715994 and the priority program SFB925. Eryin Wang received funding from the Alexander von Humboldt Foundation. We thank Rashmi Singla for her help on the optical setup. We thank Michael Volkmann, Elena König, and Peter Licht for their technical assistance. We are also grateful to Benedikt Schulte, Boris Fiedler, and Birger Höhling for their support in the fabrication of the electronic devices used on the measurement setup, and to Jörg Harms for assistance with graphics.

# Probing photo-induced granular superconductivity in $K_3C_{60}$ thin films with an ultrafast on-chip voltmeter

J. D. Adelinia[1†], E. Wang[1†], M. Chavez-Cervantes[1], T. Matsuyama[1], M. Fechner[1], M. Buzzi[1], G. Meier[1], A. Cavalleri[1,2]

[1] Max Planck Institute for the Structure and Dynamics of Matter, Hamburg, Germany
[2] Department of Physics, Clarendon Laboratory, University of Oxford, Oxford, United Kingdom
† These authors contributed equally to this work

# Supplementary information

## Contents

# S1. Experimental setup

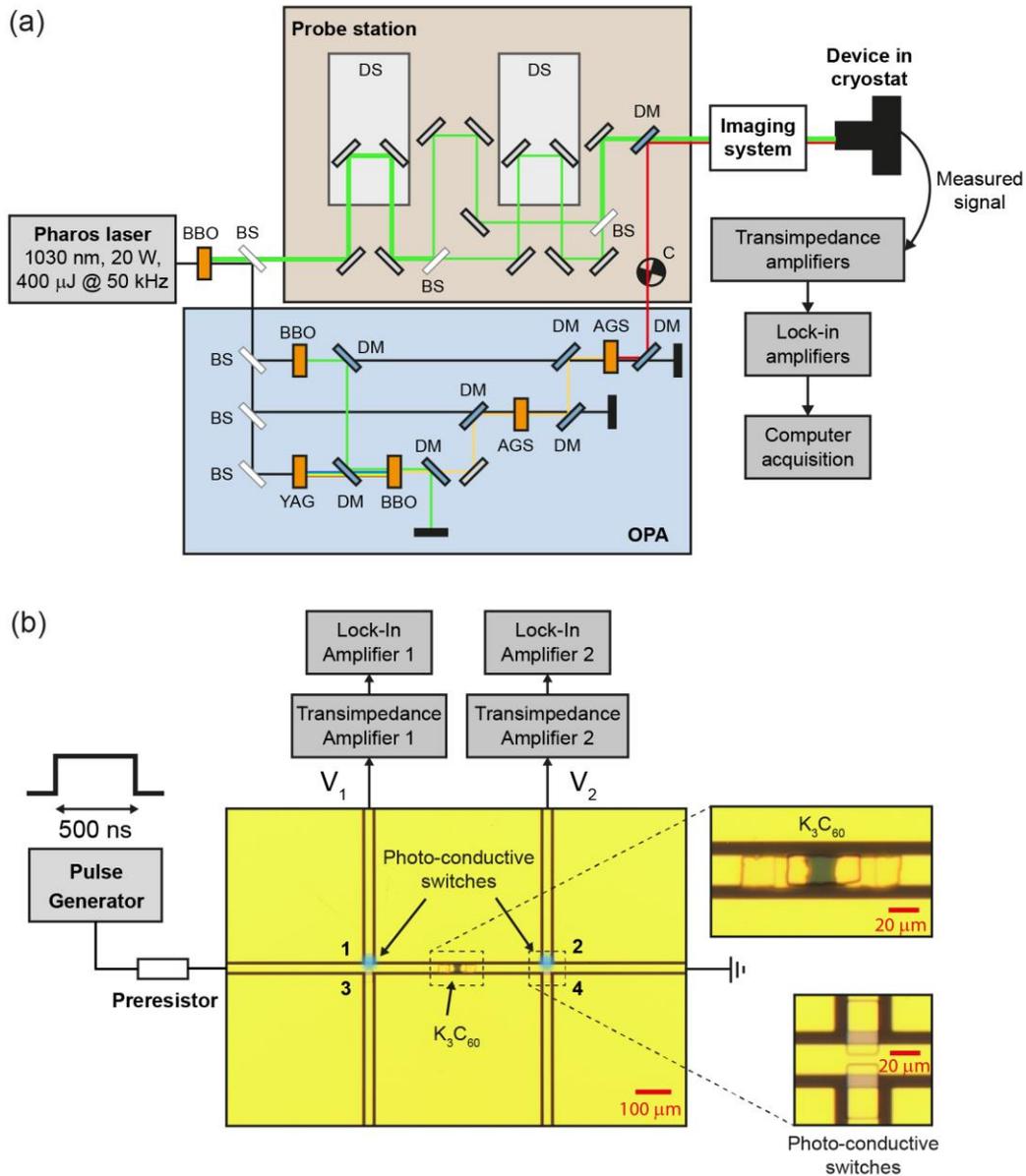

**Figure S1 | Experimental setup. a,** Optical part of the setup. Optical components are abbreviated as follows: AGS – silver thiogallate crystal, BBO – beta barium borate crystal, BS – beam splitter, C – mechanical chopper, DM – dichroic mirror, DS – delay stage, OPA – optical parametric amplifier, YAG – yttrium aluminium garnet. The wavelength of the optical pulses is represented by the colour: 1030 nm (black), 515 nm (green), 1200 nm (yellow), and 7 μm (red). **b,** Optical microscope image of the experimental device, showing connections to the external electronics.

A diagram of the experimental setup is displayed in Fig. S1. The optical part of the setup is based on a Pharos laser, which outputs pulses of 250-fs duration and 400-μJ pulse energy, with a wavelength centred at 1030 nm, at 50-kHz repetition rate. Part of the pulse energy is converted to 515-nm wavelength via second harmonic generation. This was used to operate the photo-conductive switches. The remainder serves as the input for an optical parametric amplifier (OPA) based on silver thiogallate (AGS)[33], which outputs mid-infrared pulses centred at 7-μm wavelength (43 THz) with 1-THz bandwidth. These were used to excite the $K_3C_{60}$ thin film. The mid-infrared pulses were chopped at a frequency of ~1 kHz. Optical delay lines were used to scan the mutual time delay between the 515-nm pulses and the mid-infrared pulses.

Transient currents in each detection line were sent into in-house custom-built transimpedance amplifiers with a flat frequency dependence, an optimised −3-dB point at 3 kHz, and an amplification factor of $2 \times 10^9$ V/A. The voltage output of the transimpedance amplifiers were then sent into lock-in amplifiers, which were trigged at the chopping frequency of the mid-infrared pulses. Data from the lock-in amplifier was acquired digitally using custom software written in LabVIEW.

The wavelength of the mid-infrared pulses was characterised using Fourier-transform infrared (FTIR) spectroscopy. The spectrum is displayed in Fig. S2. The spot size of the mid-infrared beam in the device plane was determined using knife-edge measurements, and is displayed in Fig. S3.

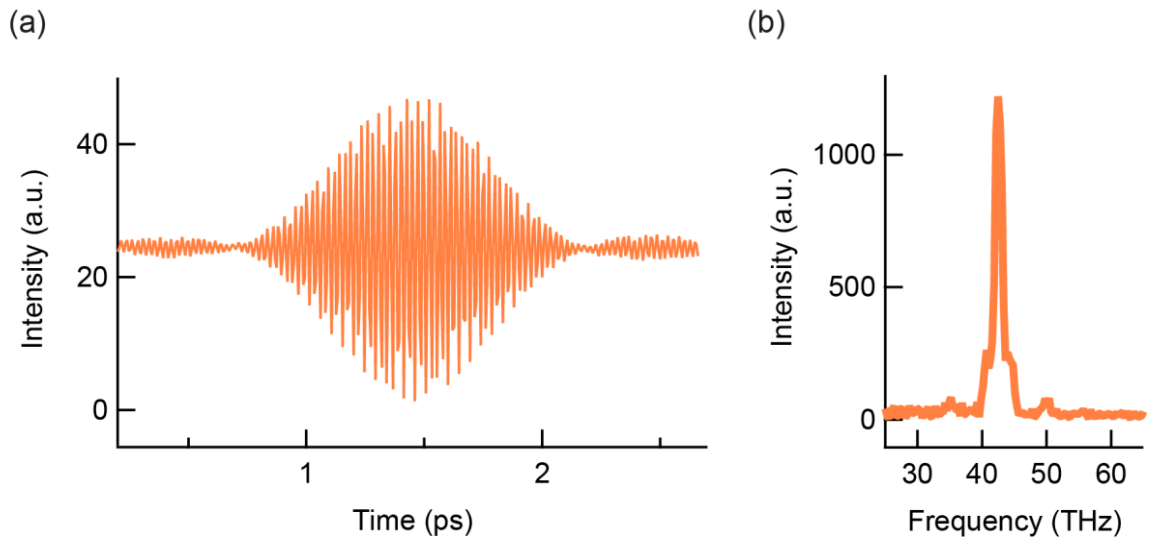

**Figure S2 | FTIR spectrum of the mid-infrared pulses. a,** Interference pattern in the time domain. **b,** Fourier transform of the data in (a).

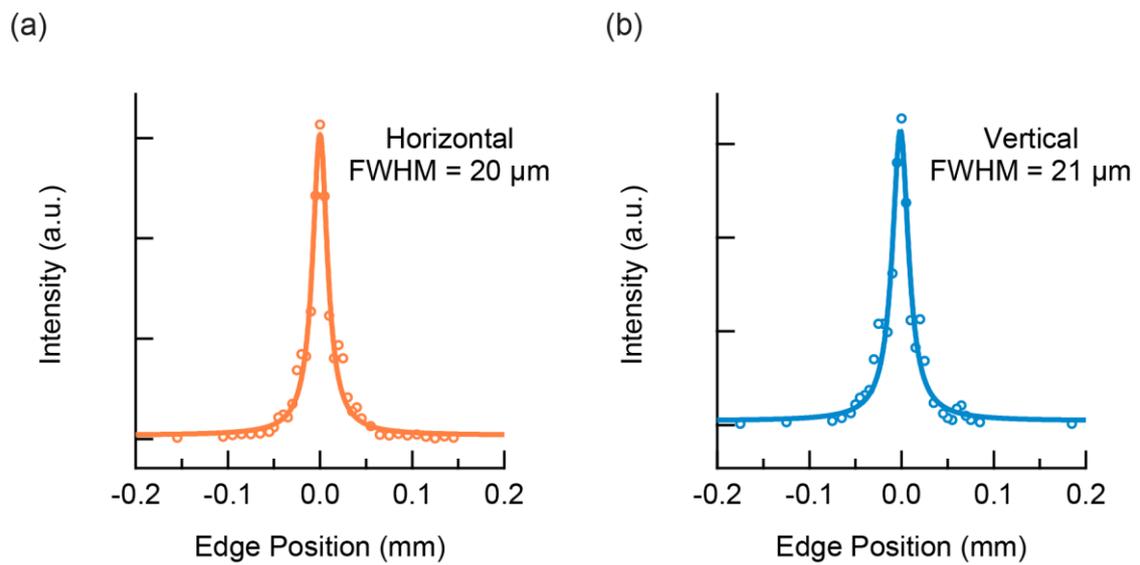

**Figure S3 | Knife-edge measurements of the mid-infrared spot size. a,** Intensity vs horizontal position of the edge. **b,** Intensity vs vertical position of the edge.

# S2. Device fabrication process

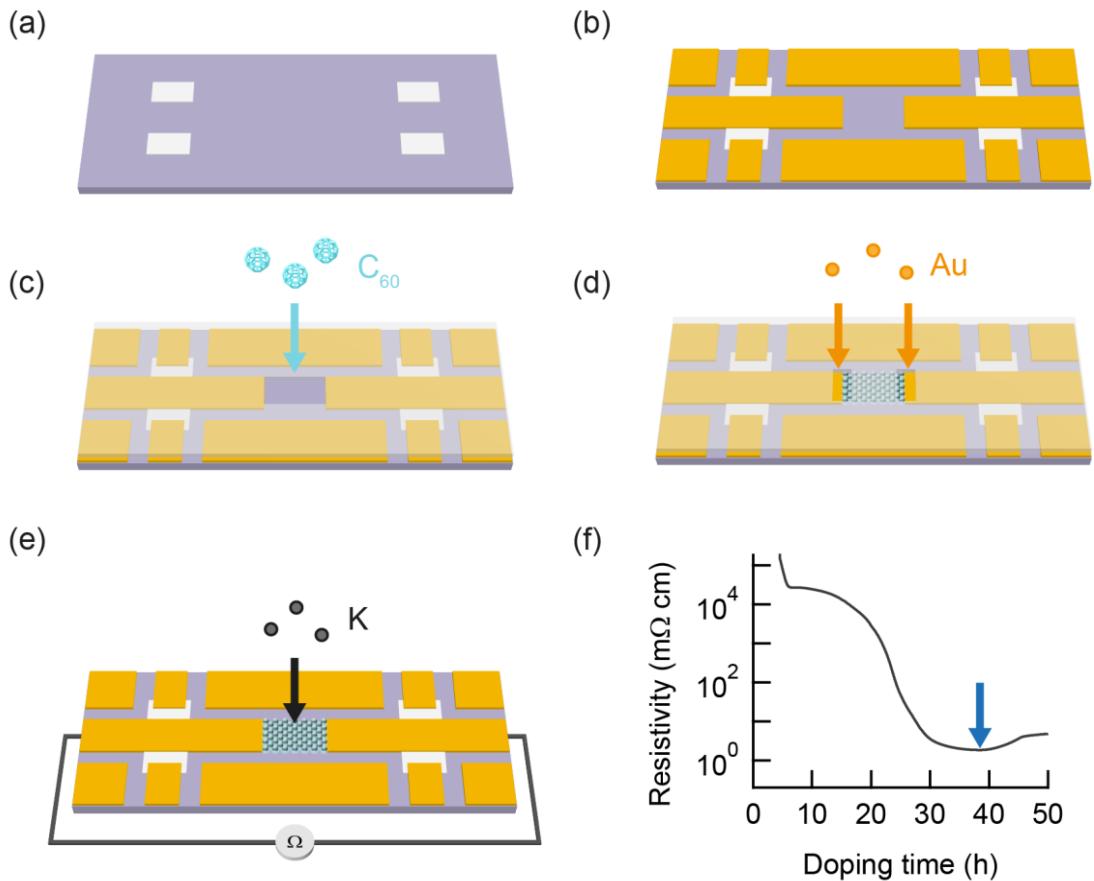

**Figure S4 | Device fabrication process.** **a-e,** Lithography and deposition steps. **f,** Resistivity of a $K_xC_{60}$ sample during potassium doping. The arrow indicates optimal doping at the resistivity minimum.

---

An illustration of the device fabrication process is displayed in Fig. S4. The photo-conductive switches and the coplanar waveguide were fabricated on a sapphire substrate using a laser lithography and electron-beam evaporation process. For the lithography steps, a two-layer photoresist was used. LOR 7B formed the bottom layer and ma-P 1205 formed the light-sensitive top layer. The lithography was carried out using a µPG 101 laser writer, and the exposed photoresist was developed in ma-D 331/S. First, patches were written and developed in the photoresist and a 200-nm layer of silicon was deposited to form the photo-conductive switches (Fig. S4a). The lift-off process was then carried out by using Remover PG to dissolve the photoresist overnight. Next, the waveguide structure was written and 10 nm of titanium was deposited, followed by 280 nm of gold (Fig. S4b), after which an identical lift-off process was conducted.

The next step was the deposition of $C_{60}$ into the waveguide. $C_{60}$ dissolves in the organic chemicals used in the photo-lithography process, so a disc of 150-μm thick sapphire was used for the shadow mask instead (Fig. S4c). The disc had a 20-μm x 30-μm hole in the centre, which was aligned to the middle of the device under an optical microscope using micrometre translation stages, and held in place using silver epoxy. The construction was transferred to a molecular beam epitaxy (MBE) chamber and degassed overnight at 300°C prior to growth. The device temperature was then reduced to 200°C and monitored using a pyrometer for the $C_{60}$ deposition. The $C_{60}$ source temperature was 380°C, providing a growth rate of ~1 nm/min.

For the electrical contact between the sample and the waveguide, an additional evaporation step of titanium/gold was carried out using a different sapphire mask (Fig. S4d).

The $C_{60}$ thin film was then doped by depositing potassium from an effusion cell source, while monitoring the resistance of the film (Fig. S4e). The resistance of the $K_xC_{60}$ decreases upon initial doping, and reaches a minimum at $x = 3$. A typical resistivity profile as a function of doping time is shown in Fig. S4f. To obtain homogeneous stoichiometry, doping was carried out in cycles consisting of one hour of doping, followed by six hours of annealing. During the doping cycles, the device was kept at 200°C and the potassium source at 100°C. During the anneal cycles, the doping was paused and the device temperature was increased to 300°C, facilitating the diffusion of potassium within the sample.

$K_3C_{60}$ is highly sensitive to oxygen. As such, the doped thin film needed to be sealed to prevent oxidation while maintaining optical and electrical access. The sealing structure is illustrated in Fig. S5. Before the doping took place, the device was attached to a sapphire plate, and electrical feedthroughs were fabricated from a sapphire ring and Torr Seal vacuum epoxy. After doping, this construction was transferred to an argon-purged glove box in a high-vacuum suitcase, and sealed with an inner indium ring, a diamond window, and vacuum epoxy. The inner indium ring serves to prevent sample damage due to outgassing when the epoxy cures, while the epoxy itself provides a more stable seal after removal from the glove box. The device was then contacted to a printed circuit board via the electrical feedthroughs and transferred to the optical cryostat for measurements.

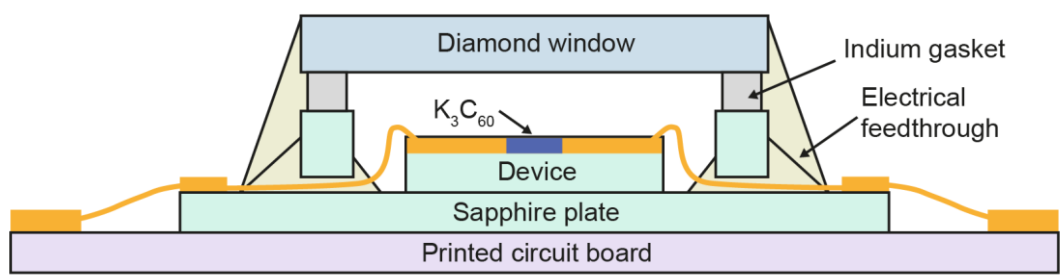

**Figure S5 | Device sealing.** Illustration of the construction used to seal the device after doping.

# S3. Device and sample characterisation

## S3.1. Determination of the quasi-DC current

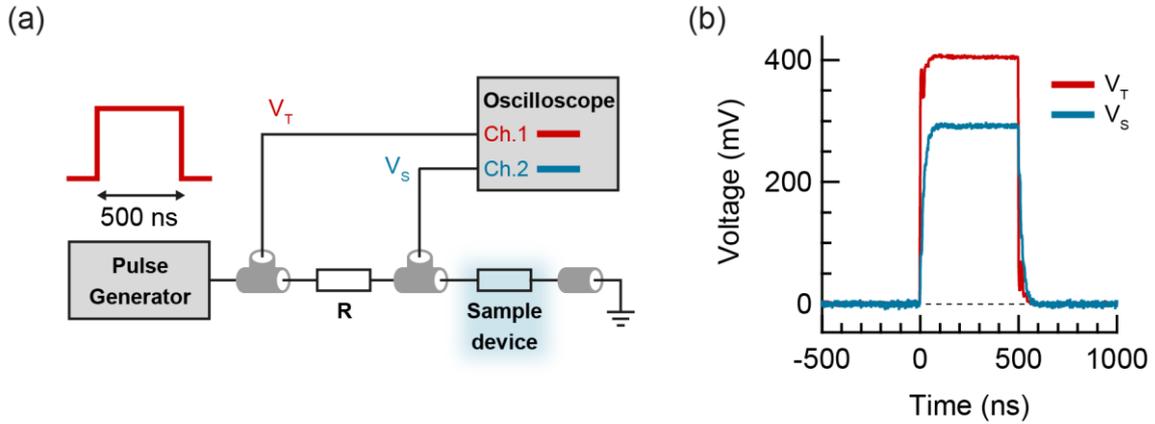

**Figure S6 | Measurement of the applied current. a,** Measurement setup with a pulsed electrical bias of 500-ns duration. **b,** Voltage drop across the sample ($V_S$) and sample + resistor ($V_T$) as a function of time for a 400-mV applied bias.

---

A quasi-DC current was applied to the $K_3C_{60}$ sample prior to photo-excitation. Flat-top voltage pulses of 500-ns duration were used for this in order to minimise current-induced heating of the sample. To determine the current flowing through the sample when the voltage pulse was present, the setup depicted in Fig. S6a was used. A known resistance R was connected in series between the voltage pulse generator and the sample device via coaxial cables. For measurements below and above T$_c$, resistances of 300 Ω and 100 Ω were used, respectively. The voltages before the resistor $V_T$ and before the sample device $V_S$ relative to ground were monitored using an oscilloscope. An example measurement for a pulsed bias of 400 mV is displayed in Fig. S6b. The stabilised voltage values could be read out at a time of 300 ns. The current flow $I$ in the device was then obtained from the equation $I = (V_T - V_S)/R$.

## S3.2. Photo-carrier lifetime in the photo-conductive switches

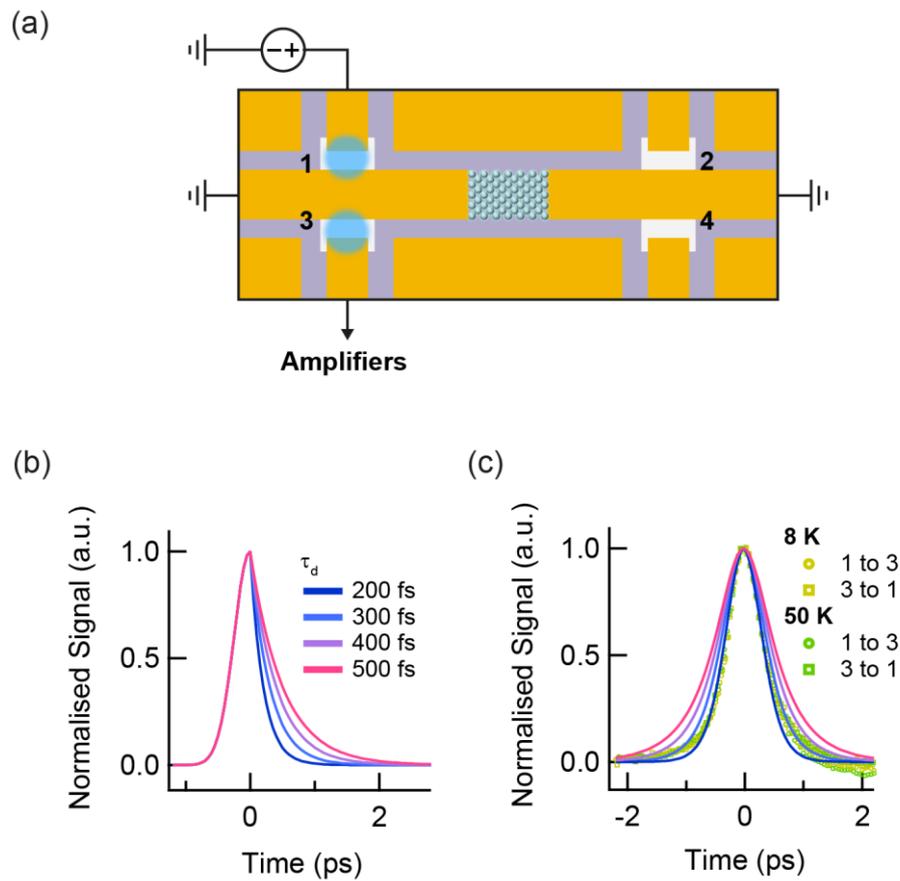

**Figure S7 | Photo-carrier lifetime measurement. a,** Measurement setup. A DC voltage bias was applied to one switch, and the launched pulses were detected at the opposite switch. **b,** Switch response functions for different exponential decay constants $\tau_d$. **c,** Autocorrelation of the switch response functions in (b), compared with the measured data when launching from different switches.

The photo-carrier lifetime in the excited photo-conductive switches is a significant factor affecting the time resolution of the experiment. To determine the lifetime, the experimental setup shown in Fig. S7a was used. Current pulses were launched from switch 1 by applying a DC voltage bias to one end and exciting the switch with a 515-nm laser pulse. The launched current pulse was then detected at switch 3. The detection switch used was directly opposite the launching switch, in order to minimise dispersion of the current pulse in the waveguide. The detected current pulse was therefore a correlation between the response function of switch 1 and that of switch 3.

The experimental data for the correlation of the two opposite switches was compared with the autocorrelation of a model switch response function for different decay times. For the model switch response, an error-function rise with a rise time of 250 fs was used, followed by an exponential decay. The rise time of 250 fs corresponds to the duration of the exciting laser pulse. Model switch response functions for different decay times are plotted in Fig. S7b. The autocorrelations of these model response functions are compared to the experimental data in Fig. S7c. The model pulse with a decay time of 300 fs provides the best fit to the data.

This approach assumes that the decay times for each switch (1 and 3) are similar, which is reasonable because the shape of the measured signal was independent of the measurement direction (circular vs square data points). The experimental data also did not change significantly with temperature.

S3.3. Waveguide characterisation

Another important factor affecting the time resolution of the experiment is the dispersion in the coplanar waveguide. To determine the impact of this, the frequency-dependent transmittance in a 200-µm long waveguide was simulated in CST studio suite. The result is shown in Fig. S8a. The transmittance drops off sharply at high frequencies, with the −3-dB point at approximately 700 GHz.

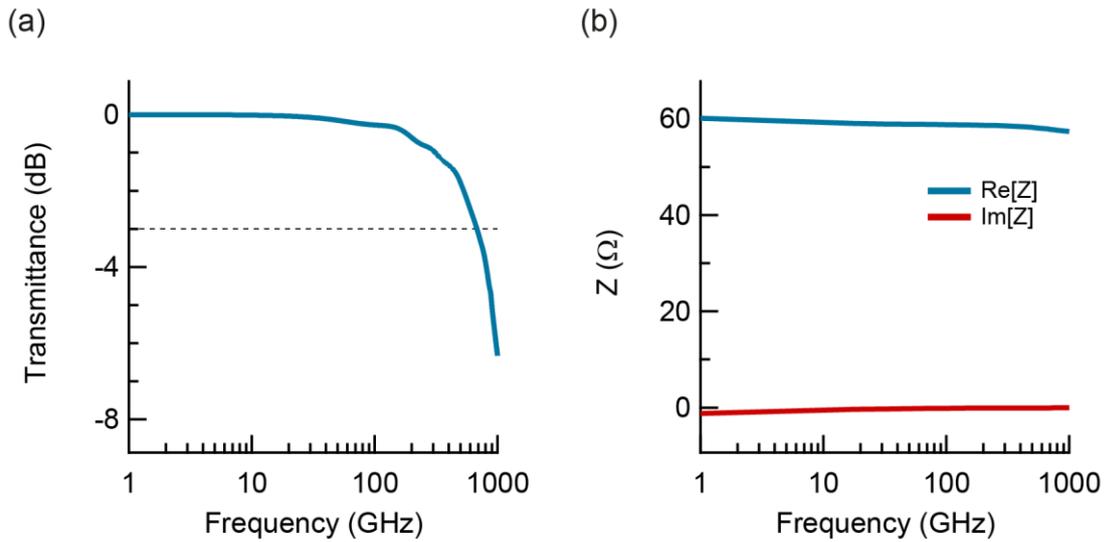

**Figure S8 | Waveguide characterisation. a,** Simulated transmittance versus frequency of a 200-μm long section of the coplanar waveguide. **b,** Real and imaginary components of the wave impedance versus frequency.

The simulated frequency-dependent complex impedance is shown in Fig. S8b. Both real and imaginary components display little variation within the experimental bandwidth. The real component Re[Z] is approximately 59 Ω, while the imaginary component Im[Z] is < 0.2 Ω and is therefore negligible.

S3.4. $K_3C_{60}$ thin film characterisation

A line profile from an AFM micrograph of the $C_{60}$ film edge is displayed in Fig. S9, showing a film thickness of approximately 100 nm.

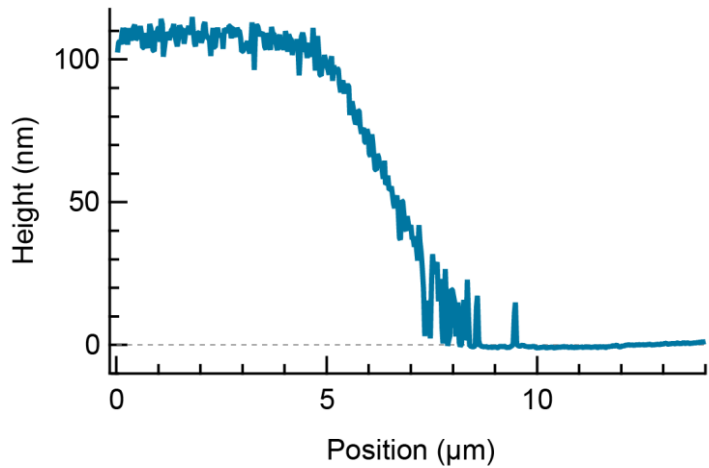

**Figure S9 | Edge of the K₃C₆₀ thin film.** Line cut of an AFM micrograph of the sample.

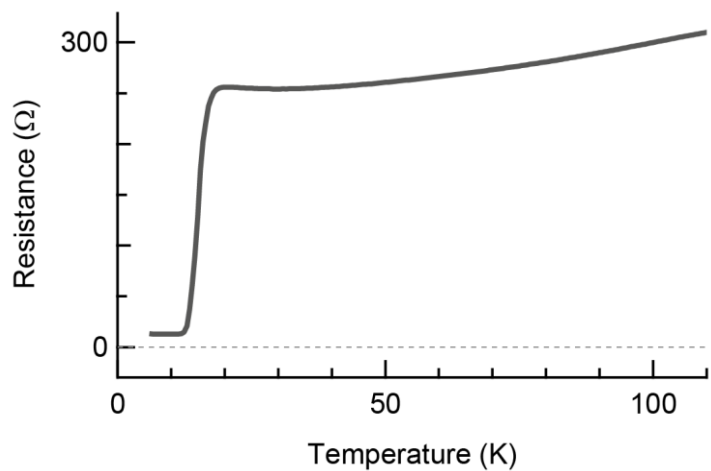

**Figure S10 | Resistance vs temperature of the K₃C₆₀ thin film.**

The resistance vs temperature of the K$_3$C$_{60}$ thin film, as measured in a two-contact geometry, is displayed in Fig. S10. The measurement displays a decrease in the sample resistance with decreasing temperature, consistent with the metallic behaviour indicated by previous studies[34,35]. The measurement was carried out with a DC current bias of 1 μA.

# S4. Data analysis

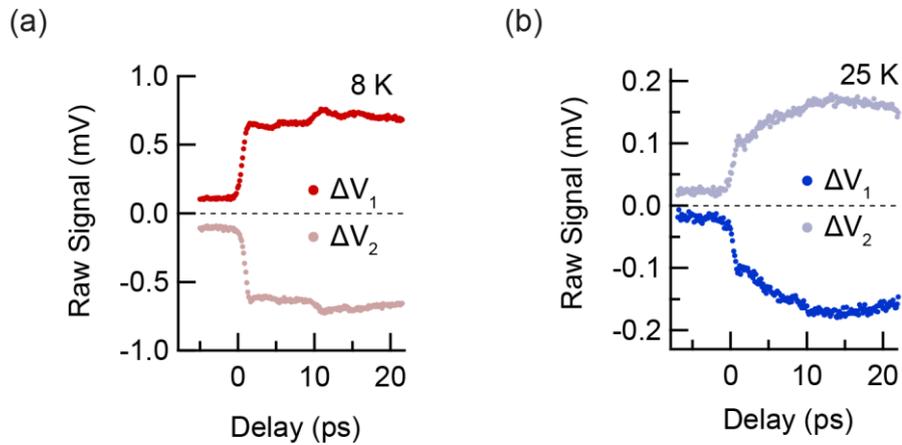

**Figure S11 | Raw transient voltage changes across the photo-excited $K_3C_{60}$ thin film. a,** Transient voltage changes measured at 8 K with an applied current of 0.3 mA. **b,** Transient voltage changes measured at 25 K with an applied current of 1.0 mA.

The raw experimental data for the temperatures 8 K and 25 K is displayed in Fig. S11. The full analysis procedure is detailed below.

## S4.1. Subtraction of 0-mA signal

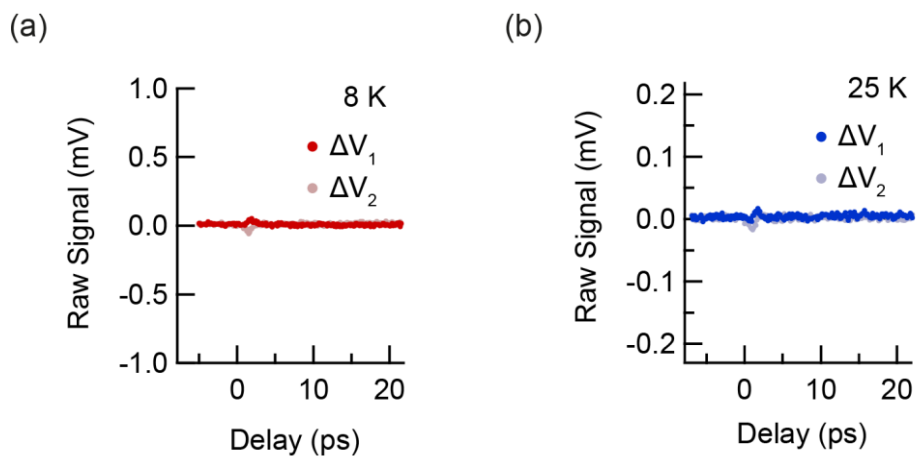

**Figure S12 | Transient voltage changes upon photo-excitation without current. a,** Zero-bias voltage changes measured at 8 K. **b,** Zero-bias voltage changes measured at 25 K.

The first step in analysis was to subtract the signal measured at 0 mA bias from the measured data at finite bias. The mid-infrared pulse has finite intensity in the gap between the $K_3C_{60}$ thin film and the waveguide. As such, the excitation pulse can couple into the waveguide and induce an electrical signal, which would be detected concurrently with the photo-induced changes in voltage drop. This effect is small, but since it is also present when there is no quasi-DC current flowing in the device, it can nevertheless be isolated and subtracted. The 0-mA signals at 8 K and 25 K are displayed in Fig. S12, and were much smaller than the measured signal at finite bias.

S4.2. Subtraction of the offset

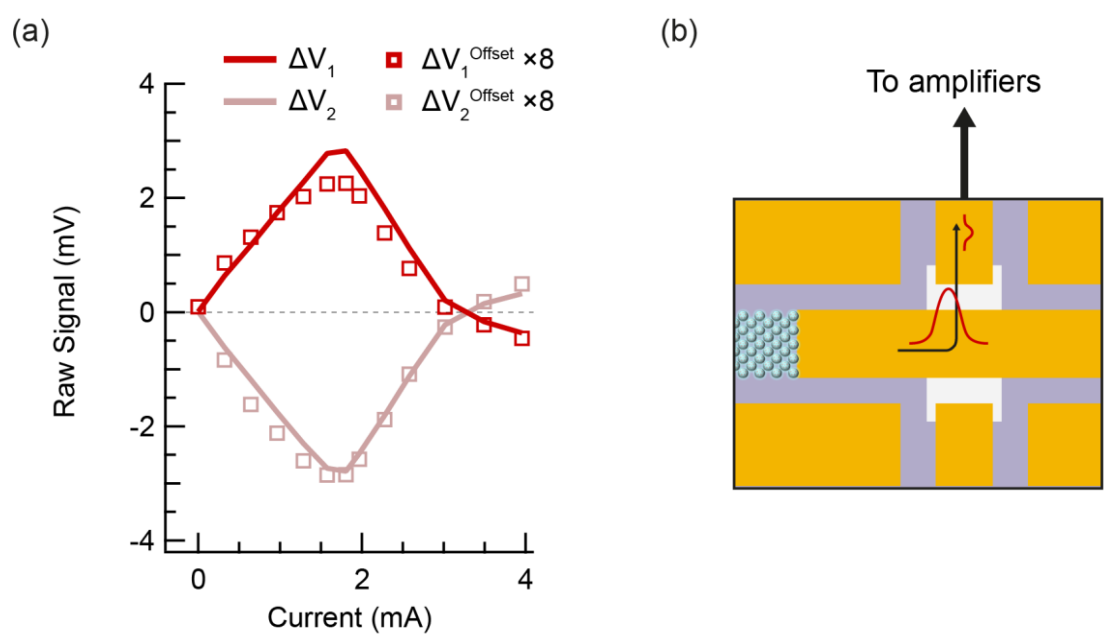

**Figure S13 | Offset before time-zero. a,** Transient voltage changes at 8 K (lines), and the offset before 0-ps delay (squares). **b,** Illustration of the leakage current through an inactive photo-conductive switch.

A constant-in-time offset was also observed in the raw data for each measurement. The offset before 0-ps delay is shown in Fig. S13a as a function of applied current for the measurements below $T_c$, alongside the voltage at 10-ps delay. The offset is proportional to the magnitude of the signal, and therefore likely results from currents flowing into the detection line as leakage currents (Fig. S13b). This results in a voltage at the amplifier that is independent of the arrival time of the switch trigger pulse. A constant offset was therefore subtracted from the data to compensate for this.

S4.3. Subtraction of reflections

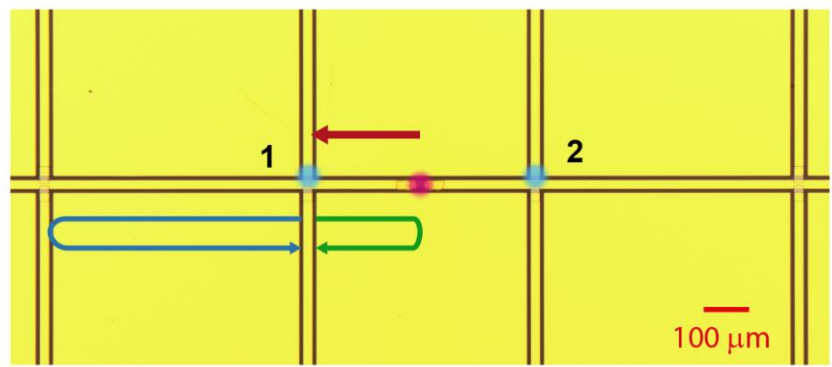

**Figure S14 | Sources of reflections.** A zoomed-out microscope image of the device. The transient voltage changes propagate towards the detection switches (red arrow), after which reflections from the sample (green arrow) and from an additional pair of switches (blue arrow) occur.

Reflections were subsequently subtracted from the resulting data. A zoomed-out optical microscope image of the device geometry is displayed in Fig. S14. Additional pairs of photo-conductive switches can be seen on the left and right sides of the image. In the experiment, a voltage pulse originated at the $K_3C_{60}$ thin film (image centre) and propagated outwards along the waveguide towards the detection switches 1 and 2.

Without loss of generality for the discussion of reflections, we will consider the leftward propagating pulse. At switch 1, the pulse is partially reflected back towards the sample and partially transmitted. The reflected pulse is again reflected from the sample and returns to switch 1 (green arrow), while the transmitted pulse is reflected from the leftmost pair of switches (blue arrow) and returns to switch 1 at a later time.

The arrival time of the reflected pulses can be estimated as follows. We consider the distance $d_1$ between switch 1 and the K$_3$C$_{60}$ thin film, the distance $d_2$ between switch 1 and the leftmost pair of switches, and the effective relative dielectric constant of the coplanar waveguide on the sapphire substrate. We have $d_1 = 280$ μm and $d_2 = 600$ μm. The effective relative dielectric constant $\epsilon_r = (\epsilon_{Al_2O_3} + 1)/2 = 5.27$, giving a pulse propagation speed $v = 0.43c$. The arrival times $t_n$ for the reflections can be calculated as

$$t_n = \frac{2d_n}{v}, \qquad (4.1)$$

giving 4.3 ps and 9.2 ps for the distances $d_1$ and $d_2$, respectively.

Copies of the measured data, scaled to 5 % and 15 % and offset in time by 4.3 ps and 9.2 ps, respectively, were subtracted. This provided a clear plateau in the data below T$_c$, and reflections were subtracted from the data above T$_c$ using the same parameters.

## S4.4. Calibration of the transient voltage changes

In order to calibrate the magnitude of the transient voltage changes, a DC voltage bias $V_b$ was applied to the photo-conductive switch as depicted in Fig. S15a. The switch was illuminated with laser pulses at 515-nm wavelength, which were chopped at a frequency of ~1 kHz. Upon illumination, a voltage $V_d$ was detected using the same amplification chain as in the experiment. Figs. S15b and S15c display $V_d$ versus $V_b$ for switches 1 and 2 at a device temperature of 25 K. The switch response was linear for all temperatures. Therefore, given only the detected voltage, the ratio of $V_b$ to $V_d$ provided a calibration factor $\kappa_S$ relating the measured voltage at switch $S$ to the voltage present in the waveguide.

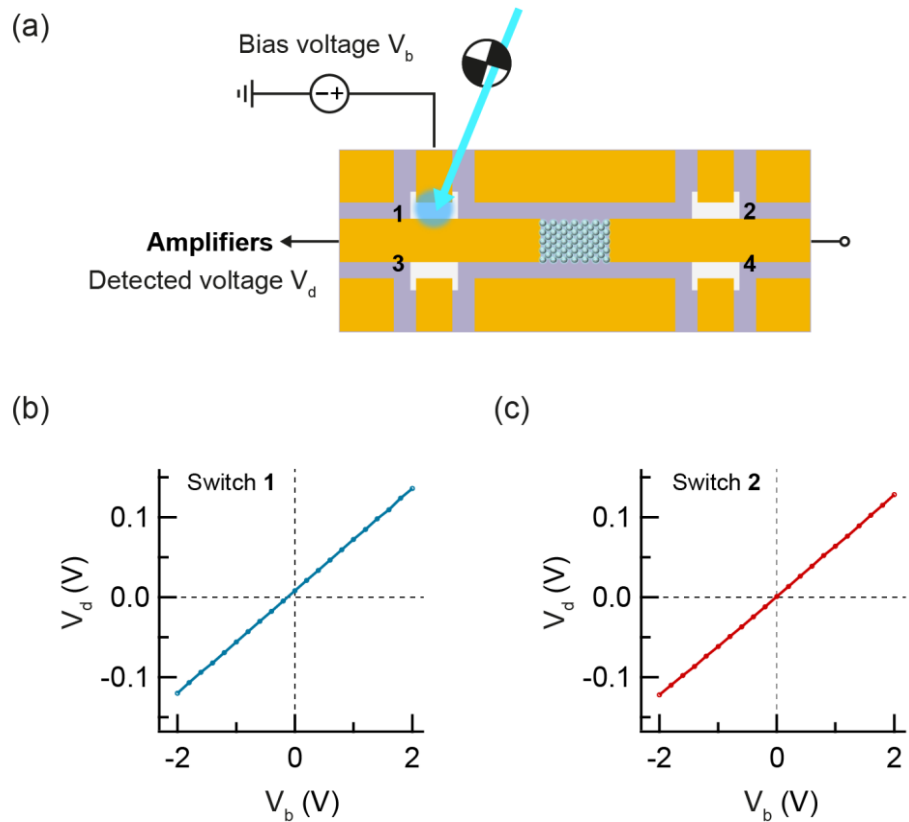

**Figure S15 | Photo-conductive switch calibration. a,** Measurement schematic. **b,** Representative calibration curve for switch 1. **c,** Representative calibration curve for switch 2.

However, directly using the $\kappa_S$ obtained from this measurement is expected to underestimate the magnitude of the voltage changes for two reasons. First, a travelling voltage pulse in the coplanar waveguide will be partially reflected from the photo-conductive switch. This can be accounted for by considering the transmittance of the switch. With the detected reflections discussed in S4.3, we see that the second reflection of 15 % magnitude (blue arrow in Fig. S14) involves the transmission from one pair of switches and the reflection from the other pair. It follows that the switch transmittance $T$ for the transient voltages in the experiment can be estimated using $T(1-T) = 0.15$. This gives $T \approx 0.8$. We can therefore account for reflections by scaling the calibration factor obtained previously by $\kappa_S \to \kappa_S/0.8 = 1.25\kappa_S$.

Second, the laser pulses incident on the switch will result in heating and an increase in DC conductivity, leading to a leakage current. This effect will last as long as the thermal dissipation time of several nanoseconds. The leakage current therefore provides an additional contribution to $V_d$ in the calibration measurement. This means that the detected voltage with a DC bias is larger than with a several-picosecond voltage bias of the same magnitude.

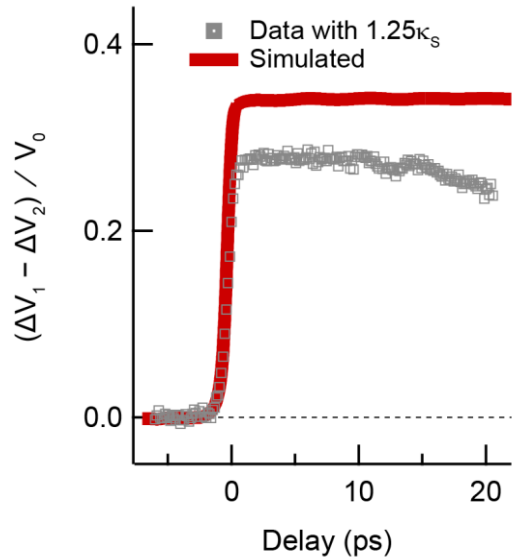

**Figure S16 | Comparison of preliminarily calibrated data with simulation.** Transient voltage changes at 8 K with a 0.3 mA bias, calibrated using a scaling factor of $1.25\kappa_S$, alongside the simulated voltage changes for disruption of superconductivity.

Fig. S16 displays the ultrafast voltmeter data from the $K_3C_{60}$ thin film at 8 K, calibrated using a calibration factor of $1.25\kappa_S$ (grey data points), alongside the simulated voltage changes for disruption of superconductivity (red line). A calibration factor of $\sim 1.5\kappa_S$ ensured consistency between the measurements at 8 K and the disruption simulation, and is consistent with a small underestimation due to leakage current in the calibration measurement. A calibration factor of $1.5\kappa_S$ was therefore used for all measurements at every temperature. Nevertheless, a systematic uncertainty exists in the magnitude of the voltage changes.

## S5. Voltage dynamics simulation

### S5.1. Voltage due to kinetic inductance for time-dependent carrier densities

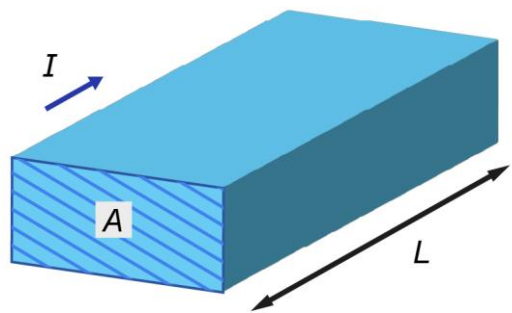

**Figure S17 | Kinetic inductance derivation.**

Kinetic inductance is an inductive electronic behaviour that arises due to the inertia of the charge carriers.[36] One way to understand this behaviour in a conductor is to consider the equations of motion of charge carriers in an electric field, neglecting scattering. We consider a material of length $L$ and cross-sectional area $A$, with an electric field $E$ applied along the direction of $L$, as illustrated in Fig. S17. For a charge carrier with charge $e$ and mass $m_e$, the equation of motion for the velocity $v$ is given by

$$\frac{dv}{dt} = -\frac{eE}{m}. \tag{5.1}$$

Assuming, for now, that the charge carrier density $n$ is constant, the current density $J$ can then be calculated as

$$\frac{dJ}{dt} = \frac{ne^2}{m_e} E. \tag{5.2}$$

We now make use of the material geometry to extract a voltage-current relation, under the assumption of a spatially uniform electric field. This gives

$$V = \left(\frac{m_e}{ne^2}\right)\left(\frac{L}{A}\right)\frac{dI}{dt}, \tag{5.3}$$

which displays equivalent behaviour to an inductor, and can be written as

$$V = L_K \frac{dI}{dt}, \tag{5.4}$$

where the kinetic inductance is therefore defined as

$$L_K = \left(\frac{m_e}{ne^2}\right)\left(\frac{L}{A}\right). \tag{5.5}$$

For a time-dependent carrier density, the current density is instead given by

$$\frac{d}{dt}\left(\frac{J}{n}\right) = \frac{e^2}{m_e} E, \tag{5.6}$$

and the voltage-current relation becomes

$$V = L_K \frac{dI}{dt} + I \frac{dL_K}{dt}. \tag{5.7}$$

S5.2. Estimation of heating due to the pump pulse

In a granular superconductor, the onset of resistivity at temperatures just below Tc is understood to result from thermal fluctuations in the phase of the order parameter, which disrupt Josephson tunnelling across the grain boundaries.[37,38] In order to consider the role this plays in the photo-induced superconducting-like state, the thermal fluctuations were estimated by accounting for the heating effect of the mid-infrared pump pulse.

First, the absorbed fraction of the mid-infrared pulse energy was estimated to be 30 % of the incident energy using a 3-layer multireflection model, where the optical properties from Ref. 34 were used. The absorbed energy $E_{Abs}$ is therefore given by $E_{Abs} = 0.3 \times Fluence \times \pi r_{Pulse}^2$, where $r_{Pulse}$ is the radius of the laser focus. To calculate the incident power, a Gaussian laser pulse with a full-width-half-maximum of 300 fs was scaled such that its integral was equal to $E_{Abs}$, as displayed in Fig. S18a.

To calculate the increase in carrier and lattice temperature in the K$_3$C$_{60}$ thin film, a two-temperature model with the following form was used[39,40]

$$c_e \frac{dT_e}{dt} = -\frac{c_e}{\tau_{e-p}}(T_e - T_p) + P(t) \tag{5.8}$$

$$c_p \frac{dT_p}{dt} = \frac{c_e}{\tau_{e-p}}(T_e - T_p) - \frac{c_p}{\tau_{es}}(T_p - T_0), \tag{5.9}$$

where $T_e$ and $T_p$ are the electron and phonon temperatures, respectively, $c_e$ and $c_p$ are the electron and phonon specific heat capacity, $\tau_{e-p}$ is approximately the electron-phonon scattering time, and $\tau_{es}$ is the "escape time", representing the rate of heat transfer from the thin film to the substrate.

For films of tens of nanometres in thickness, the escape time $\tau_{es}$ is several nanoseconds in duration and is not important on the timescales of this experiment.[39] A value of 3 ns was used here. Measurements in other metals have found values for the electron-phonon relaxation time $\tau_{e-p}$ of ~1 ps. In the simulations reported here, $\tau_{e-p} = 0.8$ ps provided the best fit to the data.

The value of the phononic specific heat capacity was obtained from fits to calorimetry data[41], where $c_p(T)$ was approximated as linear within the range of 0 – 300 K. For the electronic specific heat capacity $c_e(T) = \gamma T_e$, a value of $\gamma = 3.7 \times 10^{-2}$ J/kg K² was used.

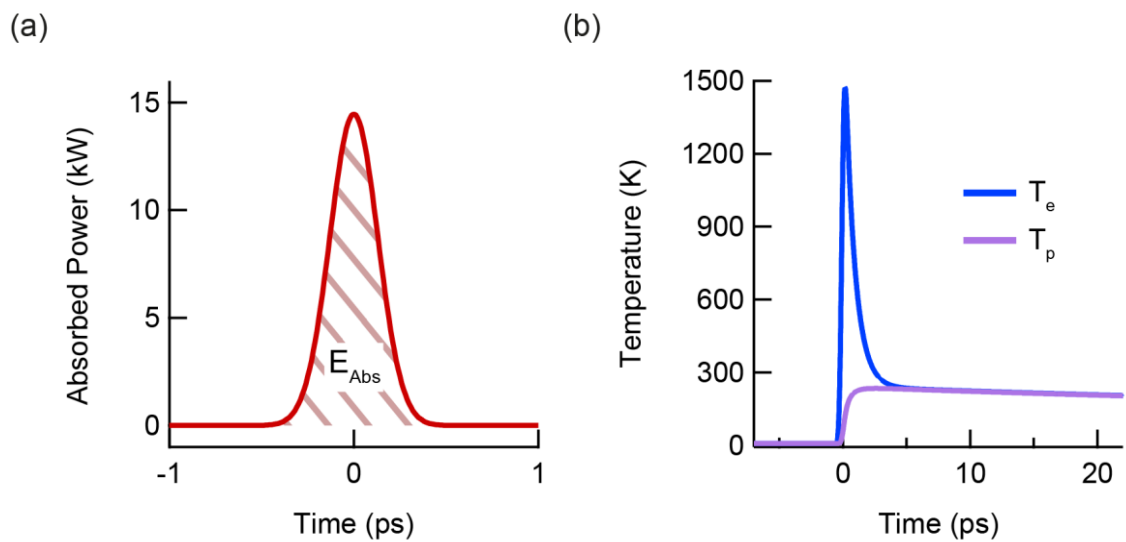

**Figure S18 | Estimation of pump-induced heating. a,** Absorbed power for a Gaussian pulse with a fluence of 5 mJ/cm². **b,** Calculated electron temperature ($T_e$) and lattice temperature ($T_p$) vs time for photo-excitation of $K_3C_{60}$ with the Gaussian pulse shown in (a).

The calculated $T_e$ and $T_p$ for a 5-mJ/cm² excitation fluence are displayed in Fig. S18b. The electron temperature rises sharply during the excitation, and the lattice temperature increases on longer timescales due to electron-phonon scattering, with the electron and lattice temperatures stabilising at ~300 K after several picoseconds. The thermal phase fluctuations in the granular two-fluid model simulation were assumed to follow the electron temperature calculated here.

S5.3. Granular two-fluid model

The full circuit model used for the simulations is displayed in Fig. S19. The granular two-fluid model, highlighted in red, was biased with a DC current. In the experiment, the transient voltage changes were detected at the photo-conductive switch positions after propagating along the coplanar waveguide, which had a wave impedance of $Z_W$. To separate the transient voltage changes in the simulation, the DC bias was isolated via low-pass filters as shown in Fig. S19.

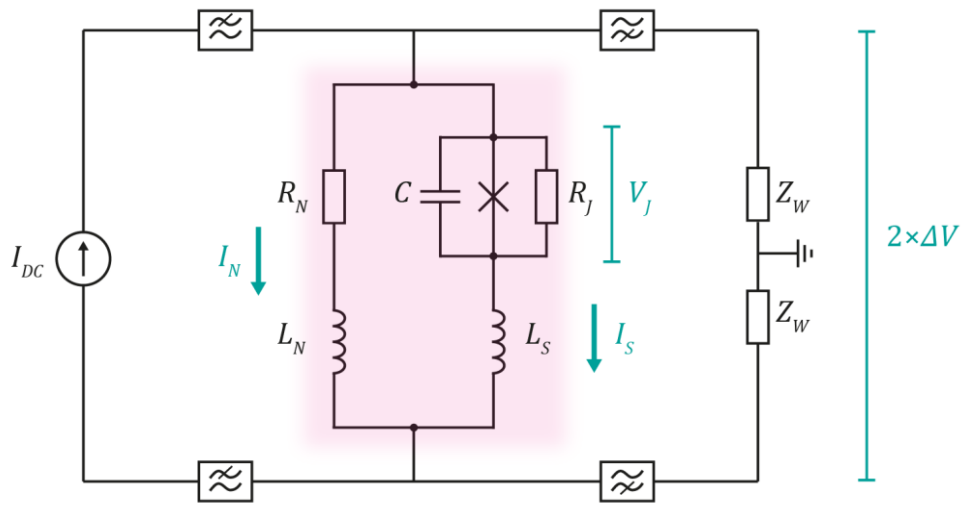

**Figure S19 | Full circuit model for granular two-fluid model simulations.** The model for the sample is highlighted in red. The right part of the circuit represents the impedance of the coplanar waveguide.

Following the treatment discussed previously[42], the granular $K_3C_{60}$ sample was modelled as an array of grains. From the atomic force microscopy measurement shown in Fig. 1 of the main text, the grain size was 100 nm × 100 nm × 20 nm, meaning the full array for the 20 μm × 20 μm × 100 nm sample was a 200 × 200 × 5 structure of grains. We can write the number of grains in each direction as $N_L = 200$, $N_W = 200$, and $N_H = 5$, where $N_L$, $N_W$, and $N_H$ represent the number of grains along the length, width, and height of the sample, respectively. The length is defined along the orientation parallel to the waveguide. The voltage changes across the length of the whole sample $\Delta V_{sample}$ could be related to the voltage changes across a single grain boundary $\Delta V_{grain}$ by

$$\Delta V_{grain} = \Delta V_{sample}/N_L . \qquad (5.10)$$

Voltage changes across the whole sample could therefore be simulated by simulating the voltage changes across a single grain boundary.

Determination of circuit parameters

The resistance $R_N$ of the normal channel was given by $R_N = R_0/f_N$, where $R_0$ is the K$_3$C$_{60}$ thin film resistance, accounting for contact and wire resistance by subtracting the measured resistance far below T$_c$, and $f_N$ is the fraction of normal carriers. Likewise, the resistance at the grain boundary $R_J$ for the superconducting state was given by $R_J = R_0/f_S$, where $f_S$ is the fraction of supercarriers. Fundamentally, $f_N + f_S = 1$. Resistances were scaled down to a single grain as $R \times (N_W \times N_H/N_L)$.

The kinetic inductance for each channel was determined in the same way, with the inductance of the normal channel given by $L_N = L_0/f_N$ and that of the superconducting channel given by $L_S = L_0/f_S$. The total kinetic inductance $L_0$ of the sample was estimated using equation (5.5). For the 20 μm × 20 μm × 100 nm thin film of K₃C₆₀, with a carrier density $n = 1.6 \times 10^{25}$ m⁻³, as extracted from optical reflectivity measurements[43], $L_0 = 22$ pH was obtained. Inductances were scaled down to a single grain as $L \times (N_W \times N_H/N_L)$.

The capacitance for a single grain boundary $C$ was estimated to be $C \approx 10^{-17}$ F, following Ref. [44].

The critical current $I_c$ of the Josephson junction was also considered to be proportional to the carrier density, following $I_c = I_c(T) \times f_S$. The temperature dependence of the critical current was given by $I_c(T) = I_c(0 \text{ K})[1 - (T/T_c)^{1.5}]$.[45] All currents, including the critical current, were also scaled down to a single grain as $I/(N_W \times N_H)$.

Finally, voltages were calculated for a single grain, and scaled up to the full sample as $V \times N_L$.

Circuit equations

The following circuit equations were derived from the granular two-fluid model:

$$2\Delta V = I_N R_N - V_0 + L_N \frac{dI_N}{dt} + I_N \frac{dL_N}{dt}, \qquad (5.11)$$

$$I_N R_N + L_N \frac{dI_N}{dt} + I_N \frac{dL_N}{dt} = V_J + L_S \frac{dI_S}{dt} + I_S \frac{dL_S}{dt}, \qquad (5.12)$$

and

$$I_{DC} - \frac{2\Delta V}{Z_W} = I_N + I_S, \qquad (5.13)$$

where $V_0$ is the initialised voltage across the sample, and other parameters are defined as displayed in Fig. S19. The voltage $V_J$ refers to the voltage drop across the resistively- and capacitively-shunted Josephson junction, and was governed by the following equations[46]:

$$\frac{dV_J}{dt} = \frac{1}{C}\left(I_S - I_C \sin\theta - \frac{V_J}{R_J} - i_n(t)\right), \tag{5.14}$$

and

$$\frac{d\theta}{dt} = \frac{2eV_J}{\hbar}. \tag{5.15}$$

The current $i_n(t)$ represents the noise current associated with the thermal phase fluctuations. A white-noise frequency spectrum was used with a zero-mean Gaussian amplitude distribution. The full-width half-maximum of the amplitude distribution was determined by comparing the equations of motion for the Josephson phase with the Langevin equation, giving $\sqrt{2k_B T/R_J}$. The circuit equations were rearranged to the form used for the numerical calculation:

$$\begin{aligned}\frac{d\Delta V}{dt} = \frac{Z_W(L_N + L_S)}{L_N L_S}\Bigg\{&-\left[2 + \frac{L_N}{Z_W(L_N + L_S)}\frac{dL_S}{dt}\right]\Delta V \\&+ \left[R_N + \frac{dL_N}{dt} - \frac{L_N}{L_N + L_S}\left(R_N + \frac{dL_N}{dt} + \frac{dL_S}{dt}\right)\right]I_N + \frac{I_{DC}L_N}{L_N + L_S}\frac{dL_S}{dt} \\&+ (V_J - V_{J0} - V_0)\Bigg\},\end{aligned} \tag{5.16}$$

$$\frac{dI_N}{dt} = \frac{1}{L_N + L_S}\left[-\frac{L_S}{Z_W}\frac{d\Delta V}{dt} - \frac{1}{Z_W}\frac{dL_S}{dt}\Delta V - \left(R_N + \frac{dL_N}{dt} + \frac{dL_S}{dt}\right)I_N + I_{DC}\frac{dL_S}{dt}\right], \tag{5.17}$$

$$\frac{dV_J}{dt} = \frac{1}{C}\left[I_{DC} - \frac{\Delta V}{Z_W} - I_C \sin\theta - \frac{V_J}{R_J} - i_n(t)\right], \tag{5.18}$$

$$\frac{d\theta}{dt} = \frac{2eV_J}{\hbar}. \qquad (5.19)$$

## S6. Additional data

S6.1. Photo-excitation below $T_c$

S6.1.1. Measurements with low excitation fluence

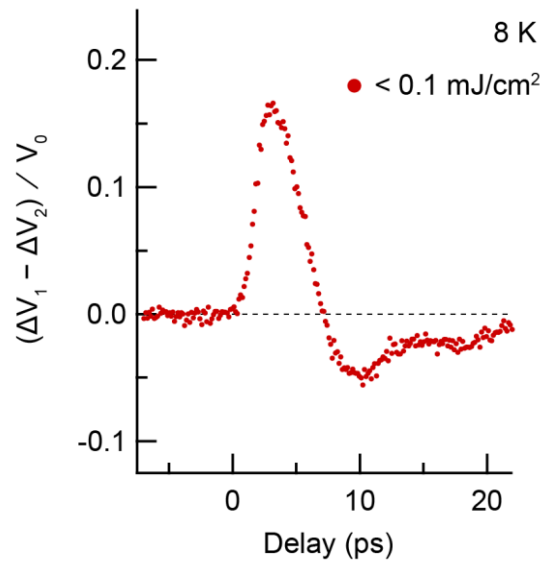

**Figure S20 | Low-fluence photo-excitation below $T_c$.** Normalised voltage changes versus pump-probe delay for photo-excitation at 8 K with a fluence of < 0.1 mJ/cm². Data were taken with a bias current of 0.3 mA.

The data for photo-excitation below $T_c$ displayed in Fig. 3 of the main text was taken with an excitation fluence of 5 mJ/cm². This resulted in a large increase in temperature of the $K_3C_{60}$ thin film and hindered the recovery of the superconducting state. The recovery of the superconducting state is expected to result in a negative voltage drop due to the decrease in kinetic inductance.[47] By photo-exciting the sample with a much smaller excitation fluence of < 0.1 mJ/cm$^2$, the heating of the sample was significantly reduced. The voltage changes upon low-fluence photo-excitation at 8 K are displayed in Fig. S20, where a negative voltage change can be seen at around 10-ps delay.

S6.1.2. Photo-excitation above the critical current

A negative change in voltage drop was also observed upon photo-excitation of the K$_3$C$_{60}$ thin film at temperatures below T$_c$, when the applied bias current was above the critical current of the equilibrium superconducting state. The time-resolved change in voltage for photo-excitation at 8 K under a 4-mA bias is displayed in Fig. S21. A two-timescale voltage response was also observed under these conditions, analogous to the voltage changes observed for the photo-induced state above T$_c$.

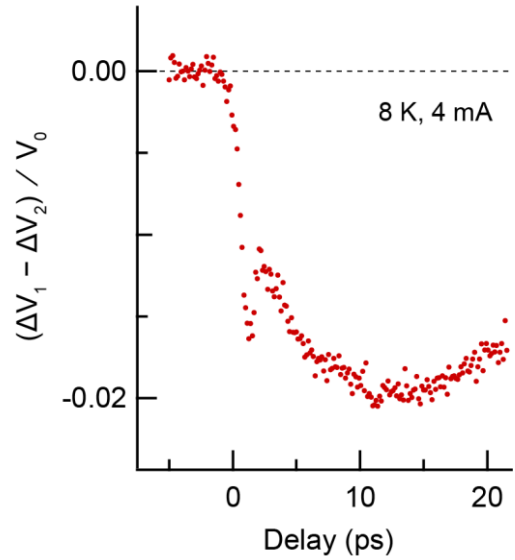

**Figure S21 | Photo-excitation above $I_c$.** Normalised voltage changes versus pump-probe delay for photo-excitation at 8 K with an applied current of 4 mA (~$2I_c$). Data were taken with a fluence of 5 mJ/cm².

S6.2. Photo-excitation above $T_c$

S6.2.1. Fluence dependence

The fluence-dependent voltage changes for photo-excitation at 25 K are shown in Fig. S22. At low fluences, the photo-induced voltage change had a shorter lifetime, and followed an exponential decay rather than the two-timescale behaviour observed at high fluences. Upon increasing the fluence, the voltage changes increased in magnitude and the spike-and-dip structure emerged, saturating at approximately 5 mJ/cm².

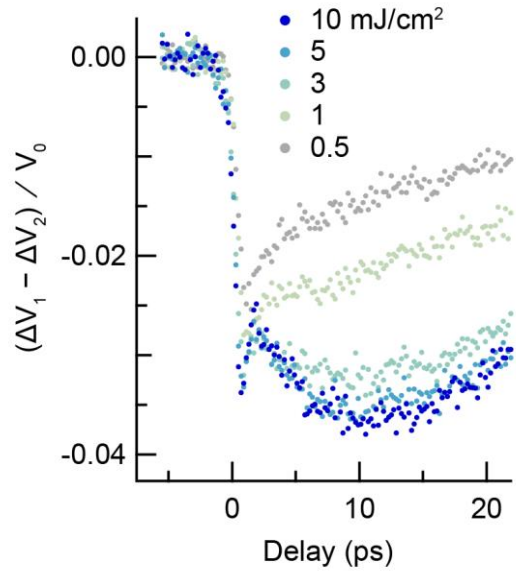

**Figure S22 | Fluence dependence for the photo-induced state above $T_c$.** Normalised voltage changes versus pump-probe delay for photo-excitation at 25 K for mid-infrared excitation fluences of 0.5, 1, 3, 5, and 10 mJ/cm². Data were taken with applied current of 1 mA.

S6.2.2. Current dependence vs fluence

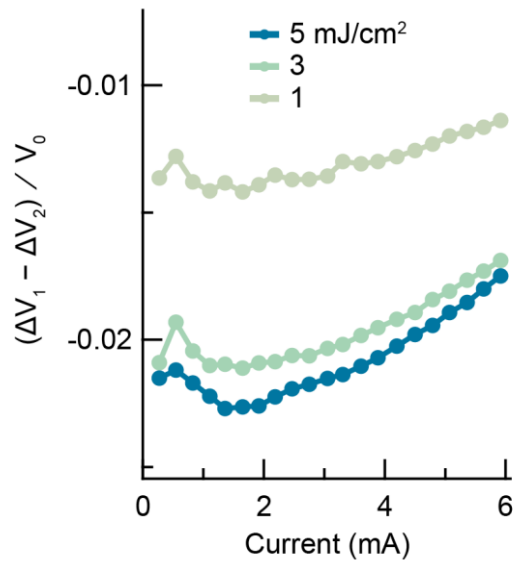

**Figure S23 | Fluence dependence of the nonlinear I-V behaviour.** Normalised voltage changes versus current for photo-excitation at 25 K, with mid-infrared excitation fluences of 1, 3, and 5 mJ/cm².

Fig. S23 displays the current dependence of the voltage changes at different fluences. Although the signal was reduced for smaller excitation fluence, a nonlinear current dependence was still observed at 1 mJ/cm².

# References

33   Kozich, V., Moguilevski, A. & Heyne, K. High energy femtosecond OPA pumped by 1030nm Yb:KGW laser. *Optics Communications* **285**, 4515-4518 (2012).

34   Degiorgi, L. *et al.* Optical properties of the alkali-metal-doped superconducting fullerenes: $K_3C_{60}$ and $Rb_3C_{60}$. *Physical Review B* **49**, 7012-7025 (1994).

35   Klein, O., Grüner, G., Huang, S. M., Wiley, J. B. & Kaner, R. B. Electrical resistivity of $K_3C_{60}$. *Physical Review B* **46**, 11247-11249 (1992).

36   Meservey, R. & Tedrow, P. M. Measurements of the Kinetic Inductance of Superconducting Linear Structures. *Journal of Applied Physics* **40**, 2028-2034 (1969).

37   England, P., Venkatesan, T., Wu, X. D. & Inam, A. Granular superconductivity in $R_1Ba_2Cu_3O_{7-\delta}$ thin films. *Physical Review B* **38**, 7125-7128 (1988).

38   Abraham, D. W., Lobb, C. J., Tinkham, M. & Klapwijk, T. M. Resistive transition in two-dimensional arrays of superconducting weak links. *Physical Review B* **26**, 5268-5271 (1982).

39   Semenov, A. D., Nebosis, R. S., Gousev, Y. P., Heusinger, M. A. & Renk, K. F. Analysis of the nonequilibrium photoresponse of superconducting films to pulsed radiation by use of a two-temperature model. *Physical Review B* **52**, 581-581 (1995).

40   Sobolewski, R. Ultrafast dynamics of nonequilibrium quasi-particles in high-temperature superconductors. *SPIE* **3481**, 480-491 (1998).

41   Allen, K. & Hellman, F. Specific heat of $C_{60}$ and $K_3C_{60}$ thin films for T=6—400 K. *Physical Review B* **60**, 11765-11772 (1999).

42   Wang, E. *et al.* Superconducting nonlinear transport in optically driven high-temperature $K_3C_{60}$. *Nature Communications* **14**, 7233 (2023).

43   Mitrano, M. *et al.* Possible light-induced superconductivity in $K_3C_{60}$ at high temperature. *Nature* **530**, 461-464 (2016).

44   McIver, J. W. *et al.* Light-induced anomalous Hall effect in graphene. *Nat Phys* **16**, 38-41 (2020).

45   Tinkham, M. *Introduction to Superconductivity, 2nd Edition*.  (Dover Publications, Inc., 2004).

46   Ambegaokar, V. & Halperin, B. I. Voltage Due to Thermal Noise in the DC Josephson Effect. *Physical Review Letters* **22**, 1364-1366 (1969).

47   Heusinger, M. A., Semenov, A. D., Nebosis, R. S., Gousev, Y. P. & Renk, K. F. Nonthermal Kinetic Inductance Photoresponse of Thin Superconducting Films. *IEEE Transactions on Applied Superconductivity* **5**, 2595-2598 (1995).